\documentclass[a4paper,11pt]{article}
\usepackage{natbib}
\usepackage[utf8]{inputenc}
\usepackage{amsmath}
\usepackage{multirow}
\usepackage{multicol}
\usepackage{mathtools}
\usepackage{amssymb}
\usepackage{booktabs}
\usepackage{bbm}
\usepackage{xurl}
\usepackage{float}
\usepackage{tabularx}
\usepackage[font=footnotesize]{caption,subcaption} 
\usepackage{graphicx}
\usepackage[affil-it]{authblk}
\usepackage{array} % bold rows in table
\usepackage{pdflscape}
\usepackage{verbatim}
\usepackage{placeins} % controlling floats
\usepackage[dvipsnames,table,xcdraw,svgnames]{xcolor}
\usepackage{pdfpages}
\usepackage{lineno} %needed to number lines
\usepackage[percent]{overpic} %to put figures on other figures
\usepackage{chngcntr} %change counter, for instance it includes \counterwithin
\usepackage{fullpage}
\newcolumntype{C}[1]{>{\centering\let\newline\\\arraybackslash\hspace{0pt}}m{#1}}

\usepackage{pdfpages}
\usepackage{xr}

\title{\vspace{-20pt}Synchronization of endogenous business cycles}

\author[a]{Marco Pangallo}
\affil[a]{CENTAI Institute, Turin 10138, Italy}

\date{\today}

\begin{document}
    \maketitle
    \footnotetext[1]{Contact: marco.pangallo@centai.eu. I thank Paul Beaudry, Doyne Farmer, Thomas Peron and Alexander Teytelboym, as well as two anonymous reviewers, for useful comments at various stages of this work. I also thank seminar participants at the Network Science in Economics Ninth Conference (Minneapolis), European Economic Association (EEA) 2020, Computing in Economics and Finance (CEF) 2023, Progress on Difference Equations (PODE) 2023, Network Science Conference (NETSCI) 2020, Conference of Complex Systems (CCS) 2022, EconophysiX Chair, Utrecht and Sant'Anna Pisa. Most of this work is included in two chapters of my DPhil thesis at the Mathematical Institute of the University of Oxford, and was performed while I was also affiliated to the Institute for New Economic Thinking at the Oxford Martin School. I acknowledge funding from INET, Baillie Gifford and from the James S. Mc Donnell Foundation. Replication files: \url{https://zenodo.org/records/8003502}.}
    \abstract{
Business cycles tend to comove across countries. However, standard models that attribute comovement to propagation of exogenous shocks struggle to generate a level of comovement that is as high as in the data. In this paper, we consider models that produce business cycles endogenously, through some form of non-linear dynamics---limit cycles or chaos. These models generate stronger comovement, because they combine shock propagation with synchronization of endogenous dynamics. In particular, we study a demand-driven reduced-form model in which business cycles emerge from strategic complementarities within countries, synchronizing their oscillations through international trade linkages. We develop an eigendecomposition that explores the interplay between non-linear dynamics, shock propagation and network structure, and use this theory to understand the mechanisms of synchronization. Next, we calibrate the model to data on 24 countries and show that the empirical level of comovement can only be matched by combining endogenous business cycles with exogenous shocks. Despite the limitations of using a stylized model, our results support the hypothesis that business cycles are at least in part caused by underlying non-linear dynamics.}

\vspace{30pt}

\textbf{JEL Class.:} C61 (Dynamic Analysis), E32 (Business Fluctuations, Cycles), F44 (International Business Cycles)
\newpage
\renewcommand*{\thefootnote}{\arabic{footnote}}

\section{Introduction}
\label{sec:Introduction}

At one extreme, business cycles have been explained as the consequence of exogenous events that originate outside the economy, such as political decisions, natural catastrophes or wars. According to this view, economies would live in stable stationary states (up to a long-term growth trend), but their dynamics are perturbed by random shocks that pull them out of their steady state. At the other extreme, business cycles have been explained as the consequence of forces that are endogenous to the economy, such as debt dynamics, overinvestment and exuberant expectations. Under this endogenous view, the steady state of the economy is fundamentally unstable, and macroeconomic fluctuations are mathematically described by some form of non-linear dynamics---limit cycles or chaos. Of course, it is possible to combine these two extreme views by adding exogenous shocks on top of endogenous macroeconomic dynamics. There are fundamental scientific and policy implications to understanding the contribution of exogenous vs. endogenous forces, and so this question has received enormous empirical and theoretical attention.

In this paper, we attack this problem from a new angle, taking advantage of the fact that business cycles synchronize differently if they are endogenous or exogenous. We consider a system composed of different countries that produce business cycles independently but are connected through international trade linkages. Countries are hit by idiosyncratic shocks. In this situation, exogenous business cycle models lead to comovement, or positive correlation, among countries' economic activity exclusively from the propagation of the shocks. For example, when two countries are in a steady state and country 1 is hit by a positive shock, country 1 increases its demand to country 2 and makes country 2's economy grow above the steady state, too. This produces a positive correlation between the macroeconomic dynamics of the two countries. According to the endogenous view, instead, comovement arises from the \textit{synchronization} of the non-linear dynamics of the two countries. 

Synchronization is a generic property of interacting components of a dynamical system to align their non-linear dynamics in a way that they operate in synchrony, provided they are sufficiently ``close''. (In this paper, we use the term synchronization in this technical sense, \textit{not} as a synonym of comovement and positive correlation as usual in economics.) Synchronization is a fascinating phenomenon that applies to very diverse systems such as oscillating pendula, flashing fireflies, firing neurons, and applauding audiences \citep{strogatz2004sync}. It has rarely been applied to economics.

To the extent that shock propagation and synchronization produce different empirical predictions on comovement, we can indirectly test whether the exogenous or endogenous views are a better description of the business cycle.

We find that the propagation of exogenous shocks produces a much lower comovement than the synchronization of endogenous business cycles. In particular, shock propagation falls short of generating a level of comovement that is as high as in the data, as in standard models of trade and exogenous shocks \citep{kose2006can}. By contrast, the empirical level of comovement can be matched by combining the synchronization of endogenous cycles with (relatively small) exogenous shocks. 

\begin{figure*}
\centering
\includegraphics[width=0.8\textwidth]{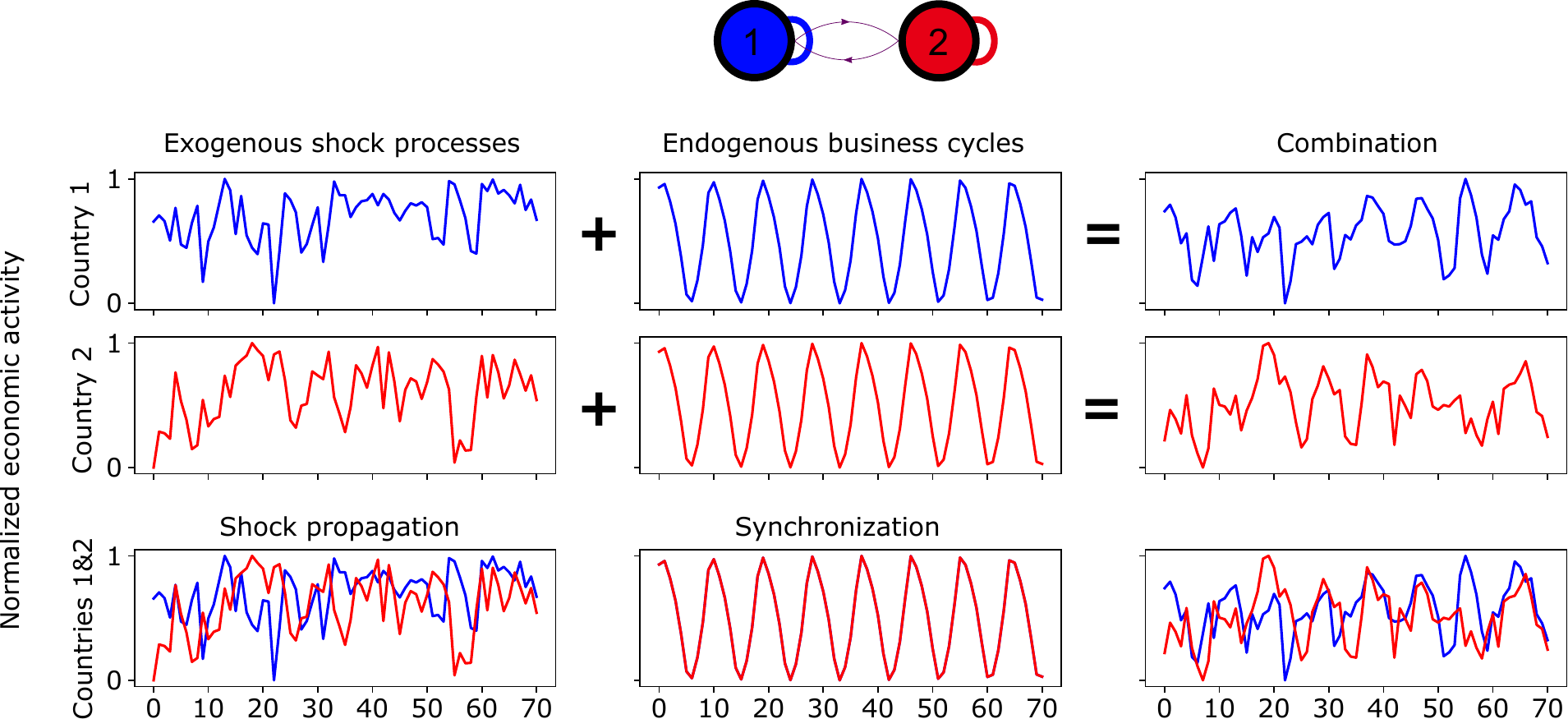}
\caption{Illustration of the main idea of the paper. Top: We consider two nodes (countries) whose economic activity depends 90\% on domestic demand (self-loops in the network representation) and 10\% on  demand by the other country (arrows). Bottom: Example dynamics of an index of economic activity for countries 1 and 2, obtained by running representative simulations for 70 time steps. Left: Exogenous shock processes. In this case, comovement originates from the propagation of these exogenous shocks. Center: Endogenous business cycles. Here, comovement originates from the synchronization of the endogenous cycles. Right: Combination of the shocks and cycles.}
\label{fig:schematic}
\end{figure*}

The economic intuition for why synchronization can help generate higher comovement is given in Figure \ref{fig:schematic}. In this illustration, we consider two countries whose economic activity depends much more on their domestic demand than on demand by the other country, as is mostly the case empirically. Under exogenous business cycles, each country is hit by an idiosyncratic shock process and produces a business cycle independently of the other country. As we can see, the correlation of the macroeconomic dynamics of the two countries is virtually zero. Intuitively, this is because shocks in one country get transmitted to the other country only by the trade linkage, which is small relative to domestic demand. On the contrary, if the two countries follow a deterministic endogenous business cycle (in this case, a limit cycle), their non-linear dynamics perfectly synchronize. Combining the exogenous shock processes with the endogenous business cycles gives a Pearson correlation of 0.29 between the dynamics of the two countries, in line with empirical observations (Section \ref{sec:tsdata}). 

These results can be interpreted as follows. The combination of persistent shocks and deterministic oscillatory dynamics implies that, at any given time, recessions occur with a certain \textit{probability}. For example, thanks to a series of positive shocks, a recession can be delayed even after a peak of the underlying limit cycle has been reached, but recession probability increases as deterministic dynamics move towards the next trough of the limit cycle. Because of synchronization, the two countries reach the peak of their deterministic dynamics roughly at the same time. The first country may fall into a recession before the second country because it is hit by a negative shock, while the second country may be enjoying a period of positive shocks. However, the second country is already predisposed to start a recession soon, and so a reduction in international demand of the first country easily drags the second country into a recession, too.

Our goal in this paper is to illustrate how synchronization theory can be useful to explain business cycle comovement. However, the theory that we develop can be applied to explain comovement of any other economic or financial time series, such as commodity, housing and stock prices. To highlight the generality of our approach, we avoid choosing models that focus on specific economic forces causing endogenous business cycles. At the same time, we avoid choosing a detailed trade model, e.g. based on gravity equations. To us, trade is just an example of an economic linkage, but other linkages could be relevant \citep{de2008will}. We instead adapt the reduced-form model proposed in \cite{beaudry2018putting}, in which endogenous cycles and connections have more abstract nature. In particular, business cycles are caused by strategic complementarities, that is, the tendency of agents to increase their action if other agents increase their action, too.  In macroeconomics, this can be thought of as the tendency of firms to increase production if their customers increase production themselves, or the tendency of households to increase consumption if other households in their social network increase consumption \citep{cooper1988coordinating}. This abstract framework can encompass many causes for endogenous business cycles.

Our first set of results is theoretical. As we want to understand the interplay between synchronization of deterministic non-liner dynamics, exogenous shocks and network structure, we adapt an approach first developed in physics, known as complete synchronization theory \citep{pecora1998master}, to our case. Here, links tend to align the non-linear dynamics of the nodes, but idiosyncratic shock processes tend to pull them apart. Complete synchronization is an elegant mathematical formalism that makes it possible to quantify the relative strength of synchronizing and desynchronizing forces, through an eigenvalue-eigenvector decomposition that takes into account the network connecting the different components. The eigenvalues and eigenvectors give information about how quickly the system may synchronize after some components are hit by idiosyncratic shocks, about which network components have more synchronized dynamics depending on the presence of clusters, etc.

Next, we specify the abstract model so that the nodes are 24 countries and a Rest of the World aggregate, and weighted, directed, links represent international demand and domestic demand (self-loops). In our empirical application, we consider different parameterizations under which the deterministic model either produces endogenous business cycles or converges to a stable steady state. We then add idiosyncratic exogenous shocks to the dynamics. We show that only a combination of endogenous business cycles and exogenous shocks can match the level of comovement that can be found in the data. Finally, we show how a few predictions of complete synchronization theory have empirical relevance.

\section{Literature review}
\label{sec:litreview}

Our paper relates first of all to the few papers that in the past have seen the appeal of synchronization theory to explain business cycles comovement \citep{haxholdt1995mode,selover1999mode,brenner2002international,
matsuyama2014globalization}. Our paper is different in several respects. The most important difference is that we use another concept of synchronization that makes it possible to compare deterministic endogenous dynamics and random shocks. Previous work used the most well-known concept of \textit{phase synchronization}, that implies that each component of the synchronizing dynamical system has a relatively well-defined natural frequency and phase; under synchronization, the frequencies would align to a common frequency and the phases would lock. The problem with this concept is that in economics it is hard to characterize oscillations with a well-defined frequency and phase. In this paper, we use the more general concept of complete synchronization. To our knowledge, this has not been applied to economics yet. The second important difference is that the papers above are purely theoretical, whereas here we test if indeed synchronization makes it possible to improve over shock propagation in explaining comovement. Other differences are that our paper (i) uses a general model that can be adapted to several economic circumstances, rather than relying on a specific business cycle model; (ii) studies synchronization on a network of an arbitrary number of agents, rather than just two agents or a trivial topology; (iii) produces ``realistic'' smooth business cycle dynamics, as opposed to e.g. limit cycles of period two.

Our paper contributes to the literature that explains international business cycles comovement. Most of this literature is empirical, trying to make sense of which channels are more likely to lead to comovement \citep{de2008will,pangallo2019endogenous}.\footnote{There are also methodologically-focused studies that use novel techniques to measure comovement, such as wavelet analysis \citep{soares2011business} and random matrix theory \citep{guerini2023synchronization}.} The channel that is unanimously found to have a positive effect on comovement is international trade \citep{frankel1998endogenity,clark2001borders, imbs2004trade, baxter2005determinants,kose2006can,calderon2007trade,inklaar2008trade,hsu2011foreign,
kleinert2015few,feldkircher2016international,di2018micro,hwang2021international}. Instead, the effect of financial integration on comovement is ambiguous. For instance, \cite{imbs2006real} find a positive effect and \cite{kalemli2013financial} find a negative effect. While it is true that financial integration helps spread financial shocks, free movement of capital implies that capital may fly from countries in distress to countries that are doing well, thus reducing comovement.\footnote{This mechanism is well-known since \cite{backus1992international}, and was recently explored in \cite{cesa2019finance} together with country-specific heterogeneity and common shocks.} Following this evidence, we consider trade as our main mechanism leading to synchronization.

Another literature attempts to explain comovement by building international real business cycle models in which comovement originates from the propagation of exogenous shocks \citep{backus1992international}. These works generally struggle to obtain a level of comovement that is as high as in the data \citep{arkolakis2009vertical,johnson2014trade,liao2015trade}, and this became known as the ``trade-comovement puzzle'' \citep{kose2006can}. As already outlined above, we contribute to this literature by showing how synchronization of non-linear dynamics is a powerful way to generate higher international comovement. 

More broadly, we contribute to the debate on the nature of business cycles as exogenous vs. endogenous. Over the last 80 years, there have been cycles of preferences towards one view or another. Following the early theoretical work on endogenous business cycles \citep{kaldor1940model,hicks1950contribution,goodwin1951nonlinear}, the real business cycles revolution \citep{kydland1982time} popularized the use of exogenous shocks to characterize business cycles. At the same time, the discovery of chaos theory revived the interest in non-linear dynamic models, and in a few years a flurry of papers on general equilibrium models producing endogenous business cycles came out \citep{boldrin1990equilibrium}. This interest died out as it was not possible to find conclusive evidence on the existence of non-linear dynamics (limit cycles or chaos) in economic time series,\footnote{\cite{barnett2000martingales,brock1986distinguishing,
barnett1988aggregation,scheinkman1989nonlinear,barnett2000martingales,
shintani2004nonparametric,hommes2006comments}.} and was only recently revived \citep{beaudry2018putting,bonart2014instabilities,asano2021emergent}. Our approach indirectly tests for non-linear dynamics in economic time series by looking at the different implications of exogenous vs. endogenous business cycles on comovement. 

%Beyond economics, our paper contributes to the literature on synchronization  \citep{pikovsky2003synchronization,arenas2008synchronization,rodrigues2016kuramoto}. The eigendecomposition that we develop here extends complete synchronization theory, which is usually concerned with the binary choice of whether complete synchronization is achieved or not. Here, we quantify how quickly dynamics at different nodes synchronizes back after shocks (if it does), in a sort of generalized impulse response function that depends on the network structure.

\section{A general framework for endogenous business cycles}
\label{sec:framework}

Over decades, researchers have proposed several economic forces that could generate endogenous business cycles.\footnote{See, among many others, the contributions and review articles in:  \cite{kaldor1940model,hicks1950contribution,goodwin1951nonlinear,goodwin1967growth,shleifer1986implementation,
boldrin1990equilibrium,foley1992contribution,silverberg1993long,bullard1994learning,matsuyama2007credit,fazzari2008cash,
de2011animal,nikolaidi2017minsky}.} Some scholars focused on the role of finance, others on real factors such as investment and inventories, still others on bounded rationality and adaptive expectations. Even more forces have been proposed, including overlapping generations effects, preference switching, search and matching, technological progress, and wage bargaining. To discuss synchronization of endogenous business cycles, it would be limiting to focus on a model that picks a particular economic force to produce business cycles in the first place, as results would appear to depend on this choice. It would instead be desirable to use a macroeconomic framework that is as general as possible.

Luckily, \cite{beaudry2015reviving} have recently proposed such a framework. The authors focus on strategic complementarities \citep{bulow1985multimarket,cooper1988coordinating} between agents as the main force leading to endogenous cycles. In their framework, agents' decisions self-reinforce through interactions, leading to an instability of the steady state, but at some point sufficiently far from the steady state, stabilizing forces set in that prevent dynamics from exploding. Many of the narratives for endogenous business cycles listed above can be cast within this framework (Section \ref{sec:modeldynamics}).

In this section, we start introducing the model in an abstract interaction network, and then consider a specification of the model in which interactions correspond to international trade. We finally discuss how this model can generate endogenous cycles.

\subsection{Abstract formulation}
\label{sec:abstractmodel}

Consider $N$ agents, indexed by $i\in\mathcal{I}=\{1,\ldots,N\}$. Denote by $x_i$ an accumulation variable representing a stock of agent $i$, and by $y_i$ a decision variable representing a flow. We can think of $y$ as production and of $x$ as inventories; we can also think of $y$ as investment or consumption of durable goods, and of $x$ as capital or net worth, respectively. 

The time evolution of the accumulation variable $x_i$ is simple. At each time step $t$, it depreciates by a factor $\delta$, and increases by the decision variable $y_{i,t}$. In formula, $x_{i,t+1}=(1-\delta)x_{i,t}+ y_{i,t}$. 

The dynamics of the decision variable $y_i$ is more involved and captures the effect of the interactions with the other agents. The planned level of $y$ for agent $i$, $y_{i,t+1}$, depends most importantly on a term $\overline{y}_{i,t}$, capturing the interactions among the decision variables $y$ of the agents. The interaction term $\overline{y}_{i,t}$ is defined as $\overline{y}_{i,t}=\sum_{j=1}^N \epsilon_{ij} y_{j,t}$, where $\epsilon_{ij}\in[0,1]$, such that $\sum_{j} \epsilon_{ij}=1$. Each term $\epsilon_{ij}$, which can be thought of as an \textit{interaction coefficient}, is the weight that the decision variable of agent $j$ has in determining the value of $\overline{y}_{i,t}$ (self-interactions $\epsilon_{ii}$ are also important). The values of $\epsilon_{ij}$, for all $i$ and $j$, define a weighted, directed, interaction network. 

The effect of $\overline{y}_{i,t}$ is mediated by a non-linear function $F\left(\cdot\right)$ that determines the effect of interactions. Suppose that the decision variables of the agents with whom agent $i$ interacts increase, so that $\overline{y}_{i,t}$ becomes larger. If, as a consequence, agent $i$'s marginal payoff goes up, one says that there are \textit{strategic complementarities} between agent $i$ and the agents with whom she interacts \citep{cooper1988coordinating}.\footnote{Letting $V_{i,t}$ be the payoff function for agent $i$ at time $t$, strategic complementarities correspond to the condition $\frac{\partial^2 V_{i,t}}{\partial y_{i,t} \partial \overline{y}_{i,t}} > 0$.} Intuitively, in case of strategic complementarities, agent $i$ decides to increase her decision variable if the agents with whom she interacts do the same, i.e. $\partial y_{i,t+1} / \partial \overline{y}_{i,t} > 0$.

We now complete the description of the model. We first assume that $y_{i,t+1}$ also linearly depends on one-step lagged values of $x$ and $y$, with coefficients $\alpha_1$ and $\alpha_2$ (further to an intercept $\alpha_0$). We take $\alpha_1$ to be negative, and $\alpha_2$ to be positive and bounded between zero and one \citep{beaudry2015reviving,beaudry2018putting}.  These assumptions reflect, respectively, decreasing returns to accumulation, i.e. willingness to avoid excessive stocks of inventories, capital or net worth, and sluggishness in the adjustment of the decision variable, i.e. difficulty to quickly modify the level of production, investment, or consumption of durable goods. The assumption on $\alpha_1$ is instrumental to avoid explosive dynamics, as it helps keeping the dynamics bounded when it wanders away from the steady state. At the same time, the assumption about $\alpha_2$ introduces realistic smoothness in the dynamics.\footnote{A problem of many endogenous business cycle models is that they generate a ``sawtooth'' dynamics that has no persistence, in stark contrast to empirical time series.} 

We finally add idiosyncratic shock terms $u_{i,t}$ to the evolution of the decision variables. We parameterize this shock process as an AR(1), i.e. $u_{i,t+1}=\rho u_{i,t} + \iota_{i,t}$, where $\rho\in[0,1]$ is a persistence parameter and $\iota_{i,t}$ is white noise, normally distributed with mean zero and standard deviation $\sigma$.

Our model for endogenous business cycles is thus fully specified by the following equations, for all agents $i$:
\begin{equation}
 \begin{array}{l}
     x_{i,t+1} = (1-\delta) x_{i,t} + y_{i,t},\\
     y_{i,t+1} = \alpha_0+\alpha_1 x_{i,t} + \alpha_2 y_{i,t} + F \left( \overline{y}_{i,t} \right) + u_{i,t},\\
     \overline{y}_{i,t}= \sum_{j=1}^N \epsilon_{ij} y_{j,t}, \hspace{10pt} \epsilon_{ij}\in[0,1], \hspace{10pt} \sum_{j} \epsilon_{ij}=1.
  \end{array}
  \label{eq:dynsyst}
\end{equation}

In the case in which all interaction coefficients $\epsilon_{ij}$ are identical and equal to $1/N$ (complete and homogenous interaction network), and all agents behave alike, $x_{i,t}=x_{j,t}$ and $y_{i,t}=y_{j,t}$, for all agents $i$ and $j$ and for all times $t$ (and shocks are identical across all agents), Eqs. \eqref{eq:dynsyst} recover the model in \cite{beaudry2015reviving}.\footnote{\label{test} A minor difference with respect to \cite{beaudry2015reviving} is that $y_{i,t+1}$ depends only on variables at $t$, while in \cite{beaudry2015reviving} it depends on both variables at $t$ and (contemporaneous) variables at $t+1$. As there did not seem to be much difference in the implied dynamics, we chose this form as it was computationally simpler. Economically, it corresponds to assuming that the decision variables of the agents with whom agent $i$ interacts have a lagged effect on her decision.}

\subsection{International model}
\label{sec:internationalmodel}

We now interpret the agents in the abstract model as countries. Following the discussion in the literature review section, we interpret the interaction coefficients as trade linkages. As we stressed already, we do not aim at building a particular microfounded trade model, but rather show how much comovement a general model with connections parameterized as trade linkages can produce under exogenous vs. endogenous business cycles.

A practical difference with respect to the abstract model described in the previous section is that countries have diverse sizes. This would imply that the accumulation and decision variables would be on different scales, making it necessary to consider a different set of parameters for each country. To avoid this impractical situation, we work with ``oscillation'' variables $x_{i,t}$ and $y_{i,t}$ that define the oscillations independently of the scale, while moving all the effects of the scale to the interaction coefficients. 

To do so, we let total investment (or production, or durable consumption) be given by $Y_{i,t}=\widetilde{Y}_i y_{i,t}$ and total capital (inventories, stock of durables) be $X_{i,t}=\widetilde{Y}_i x_{i,t}$. In these expressions, $\widetilde{Y}_i$ denotes the steady state value of $Y_{i,t}$, because with our parameter restrictions $y_{i,t}$ oscillates with mean one. Then
\begin{equation}
 \begin{array}{l}
     x_{i,t+1} := X_{i,t+1}/\widetilde{Y}_i =  (1-\delta) x_{i,t} + y_{i,t},\\
     y_{i,t+1} := Y_{i,t+1}/\widetilde{Y}_i = \alpha_0+\alpha_1 x_{i,t} + \alpha_2 y_{i,t} + F \left( \overline{y}_{i,t} \right) + u_{i,t}.
  \end{array}
  \label{eq:dynsyst2}
\end{equation}

In this application, we take the interaction term $\overline{y}_{i,t}$ to represent the oscillation of the total demand received by country $i$. Letting total demand be denoted by $D_{i,t}$, we assume that its steady state $\widetilde{D}_i$ is equal to the steady state of the flow variable $Y_{i,t}$, that is $\widetilde{D}_i=\widetilde{Y}_i$. This corresponds to assuming market clearing in the steady state: depending on the interpretation for the decision variable, it could be that total demand of intermediate and final goods equals production, or that total demand of capital goods equals investment, or that total demand of durable goods equals realized consumption of such goods. Under this assumption, we can write $D_{i,t}=\widetilde{D}_i d_{i,t}=\widetilde{Y}_i d_{i,t}$, where again $d_{i,t}$ is the oscillation of demand around its steady-state value. So, in this application $\overline{y}_{i,t}:=d_{i,t}$.

The key behavioral assumption to compute total demand is that countries demand fixed proportions of imports from other countries. Letting $\widetilde{Y}_{ij}$ be the steady-state trade flow from country $i$ to country $j$, and being $a_{ij}=\widetilde{Y}_{ij}/\widetilde{Y}_j$ the fraction of imports from country $i$ over country $j$'s output, we write the total demand that country $i$ receives as $D_{i,t}=\sum_j a_{ij} Y_{j,t}$. This means that, whether country $j$ produces more or less than the steady-state value $\widetilde{Y}_j$, it still demands the fixed proportion $a_{ij}$ of imports from $i$. Equating the two formulations for $D_{i,t}$, $D_{i,t}=\widetilde{Y}_i d_{i,t}$ and $D_{i,t}=\sum_j a_{ij} Y_{j,t}$, we can express the oscillations of demand $d_{i,t}$ in terms of the oscillations of the decision variables $y_{j,t}$, as:
\begin{equation}
d_{i,t} = \frac{1}{\widetilde{Y}_i} \sum_j \frac{\widetilde{Y}_{ij}}{\widetilde{Y}_j} \widetilde{Y}_j y_{j,t} = \sum_j \frac{\widetilde{Y}_{ij}}{\widetilde{Y}_i} y_{j,t} = \sum_j \epsilon_{ij} y_{j,t},
\end{equation} 
where the coefficients $\epsilon_{ij}=\widetilde{Y}_{ij}/\widetilde{Y}_i$ are the fraction of the output of country $i$ that is \textit{exported} to $j$. Importantly, this includes the fraction of output that remains in country $i$, $\epsilon_{ii}$, representing domestic demand. These coefficients satisfy the conditions laid out in Eq. \ref{eq:dynsyst}: $\epsilon_{ij}\in[0,1]$ and $\sum_{j} \epsilon_{ij}=1$. 

In sum, the abstract model introduced in Section \ref{sec:abstractmodel} can be interpreted as a model of business cycles of countries coupled by an international trade network. Under strategic complementarities, an increase in domestic and international demand may prompt firms in several countries to increase their decision variables, further increasing domestic and international demand. This could make the steady state unstable, and, as we discuss next, the model could produce endogenous business cycles. 

\subsection{Model dynamics}
\label{sec:modeldynamics}

We formally show under which conditions the steady state loses stability in Appendix \ref{sec:uncoupled_dyn}, where we use the strength of strategic complementarities at the steady state as a bifurcation parameter.\footnote{A possible criticism of this paper is that we are using strategic complementarities both as a way to get nonlinear dynamics and as a way of generating comovement, thus making it obvious that nonlinear dynamics corresponds to high comovement. This criticism is misplaced because using other bifurcation parameters while keeping the strength of strategic complementarities fixed leads to the same results (Appendix \ref{sec:bifurcation_no_compl}).} We also present numerical simulations representing this dynamics, with and without shocks, in Appendix \ref{sec:numerical_example}. Here, we just provide the intuition for endogenous dynamics in the abstract framework, and then discuss its economic interpretation.

Consider a country whose dynamics is purely driven by internal demand by identical agents (and thus drop the subscript $i$). Starting from a situation below the steady state, the regime of strategic complementarity makes the agents increase their decision variables $y$ when the other agents are doing the same, in a way that they overshoot the steady state. However, as the agents start to over-accumulate $x$ because the value of $y$ more than offsets the depreciation $(1-\delta)x$, the rise of $y$ slows down. This is both because the agents dislike accumulation and because, as $y$ increases, the function $F(\cdot)$ stops increasing or even decreases (Figure \ref{fig:stability_model}B in Appendix \ref{sec:uncoupled_dyn}). In this situation of large $y$, strategic complementarities are very weak, or we may even be in a regime of strategic \textit{substitutability}, where an increase of $y$ by one agent makes the other agents decrease their decision variables. Soon, $y$ reverts and starts to decrease, leading to a prolonged recession. At some point, the decrease in $y$ slows down as agents liquidated excessive stock, and we are back to a regime of strategic complementarity. The process restarts. 

Depending on the interpretation for the accumulation variable $x$, the decision variable $y$, and the function $F$, the narrative above could correspond to several economic forces causing endogenous business cycles. \cite{beaudry2018putting}, for example, consider a microfounded model in which $x$ is net worth, $y$ is consumption of durable goods, and $F$ models banks' willingness to give loans. In good times, lending is perceived safe, and so agents can borrow to consume more durable goods. This further strengthens the economic boom, making lending to be perceived even safer (strategic complementarity). When the economy slows down because of overaccumulation, lending instead starts to be perceived less safe, making banks cut back on credit (strategic substitutability). This behavior of banks can easily trigger a recession, which lasts until agents have liquidated assets in excess, at which point the cycle starts again. This narrative \citep{beaudry2018putting} is similar to many of the previously proposed narratives for finance-based endogenous business cycles \citep{nikolaidi2017minsky}. However, the strategic complementarity-substitutability framework easily lends itself to other narratives, such as ones based on overinvestment in the Keynesian tradition \citep{kaldor1940model,hicks1950contribution,goodwin1951nonlinear}, and ones based on temporal clustering of technological innovations \citep{judd1985performance,shleifer1986implementation,matsuyama1999growing}.

\section{Complete synchronization}
\label{sec:completesync}

Synchronization theory is about the alignment of dynamics of weakly-interacting self-oscillating units \citep{pikovsky2003synchronization}. It is important that units are self-oscillating, that is, they would still display some form of self-sustaining non-linear dynamics (limit cycles or chaos) if they were uncoupled. Synchronization is not about whether stable units, when coupled, would become unstable \citep{gualdi2015endogenous}; it is about whether the dynamics of unstable units, when coupled, would align. The coupling is usually ``weak'' in the sense that the dynamics of the nodes are only weakly affected by the other nodes at a given time. This characterization of synchronization theory seems ideally suited to our case study, featuring different countries, weakly coupled through international trade (Section \ref{sec:internationalmodel}), displaying endogenous business cycles on their own (Appendix \ref{sec:uncoupled_dyn}).

A large part of synchronization theory is about the alignment of the frequency and phase of the oscillations (phase synchronization). To us, this is not so relevant for economics, as economic fluctuations do not generally display clear periodicity. We instead turn to the other part of synchronization theory, that goes under the name of \textit{complete} synchronization (also known as chaotic, or full, or identical synchronization). Under this theory, different units are described by the same parameters and so, under coupling, individual periodic dynamics would completely align. However, desynchronizing forces such as chaos and noise tend to separate individual dynamics. Complete synchronization theory is an elegant mathematical formalism that makes it possible to quantify the relative strength of synchronizing and desynchronizing forces, through an eigenvalue-eigenvector decomposition that takes into account the network connecting the different units.

\subsection{Theory}
\label{sec:masterstabilityanalysismain}

We build on the master stability approach originally proposed by \cite{pecora1990synchronization, pecora1998master}. Appendix \ref{sec:completesynctheory} derives all the equations, while here we give some intuition into how the approach works.

Master stability analysis starts by assuming that all agents are in a \textit{synchronized state} $\boldsymbol s_t=(x^s_t,y^s_t)$, in which they all behave alike. We now express the dynamics of individual agents by $x_{i,t}=x^s_t+x^\xi_{i,t}$ and $y_{i,t}=y^s_t+y^\xi_{i,t}$, where $\boldsymbol \xi_{i,t} = (x^{\xi}_{i,t}, y^{\xi}_{i,t})$ denotes a small deviation from the synchronized state, for example due to idiosyncratic shocks hitting agent $i$. By performing a Taylor expansion of the dynamical equations around the synchronized state, it is possible to derive the evolution of $\boldsymbol \xi_{i,t}$, for all agents $i$. As we show in Eq. \eqref{eq:masterstab} in Appendix \ref{sec:completesynctheory}, $\boldsymbol \xi_{i,t+1}$ depends on all the deviations of the other agents $j$, $\boldsymbol \xi_{j,t}$. So the dynamics of each agent are coupled to those of all other agents. The coupling is mediated by the \textit{normalized Laplacian}, a matrix derived from the interaction network whose eigenvalues and eigenvectors give key information about the structure and dynamics of the network. 

The key idea of the master stability approach is to perform a diagonalization that uncouples the deviations from the synchronized state into orthogonal independent components $\boldsymbol \zeta_{i,t} = (x^{\zeta}_{i,t}, y^{\zeta}_{i,t})$, called \textit{eigenmodes}. These components correspond to the eigenvalues and eigenvectors of the normalized Laplacian. By numerically computing the Lyapunov exponents of the eigenmodes dynamics, for each possible eigenvalue, one derives the so-called \textit{master stability function}. It is then possible to see that eigenmodes corresponding to smaller eigenvalues have less negative Lyapunov exponents, and so take longer to return to zero after a shock (Figure \ref{fig:master_stability_function} in Appendix \ref{sec:completesynctheory}). The eigenvectors corresponding to each eigenmode relate the eigenmodes to the dynamics of the agents.

To illustrate this, consider the simplest possible case of a network of two agents (Appendix \ref{sec:two}), and assume that each agent follows limit cycle dynamics. The smallest eigenvalue of the normalized Laplacian of this simple network is always null, and the corresponding eigenvector is $(+1,+1)$. The Lyapunov exponent corresponding to this eigenvalue is zero: therefore, any shock hitting the agents does not decay in this eigenmode. Moreover, the eigenvector shows that the shock affects both agents in the same way, so this eigenmode can be interpreted as a shift of the common dynamics of the two nodes. Conversely, the second eigenvalue is positive. It corresponds to negative Lyapunov exponent, so any shock reverts to zero after a few time steps. The eigenvector is $(+1,-1)$, so this eigenmode can be interpreted as the relative shift in the dynamics between the two agents due to the shock. In this case, the master stability analysis indicates that the synchronized dynamics is stable, in the sense that agents synchronize again after a shock.

In a more general network, the eigenmodes illustrate how long different components of the network take to synchronize after a shock. Consider a network composed by six nodes, with two cliques of three nodes each, connected by a link between two of the nodes of the cliques (Appendix \ref{sec:six}). In this case, the second smallest eigenvalue is small, and the corresponding eigenvector (\textit{Fiedler vector}) is positive at the nodes in one clique, and negative at the nodes of the other clique. All other eigenvalues are much larger. Here, synchronization occurs first within cliques and last between cliques. The second eigenmode corresponds to between-clique synchronization.

In conclusion, master stability analysis is a principled method to understanding when synchronized dynamics are stable, and, if they are, how quickly the agents revert to the synchronized state after they are hit by an idiosyncratic shock. In the case of a complex network structure, the analysis quantifies how quickly different parts of the network synchronize. The approach is semi-analytical, in the sense that the relative strength of synchronizing and desynchronizing forces is obtained by numerically computing Lyapunov exponents, but once these are computed, the key insights are obtained analytically. 

%In the literature \citep{arenas2008synchronization,eroglu2017synchronisation}, researchers usually content themselves with establishing under which conditions the synchronized state is stable. Here, we zoom into the processes that lead to synchronization after a shock hits the agents.

\subsection{Application: Complete synchronization of real-world countries}
\label{sec:real}

\begin{figure*}[!h]
\centering
\includegraphics[width=1\textwidth]{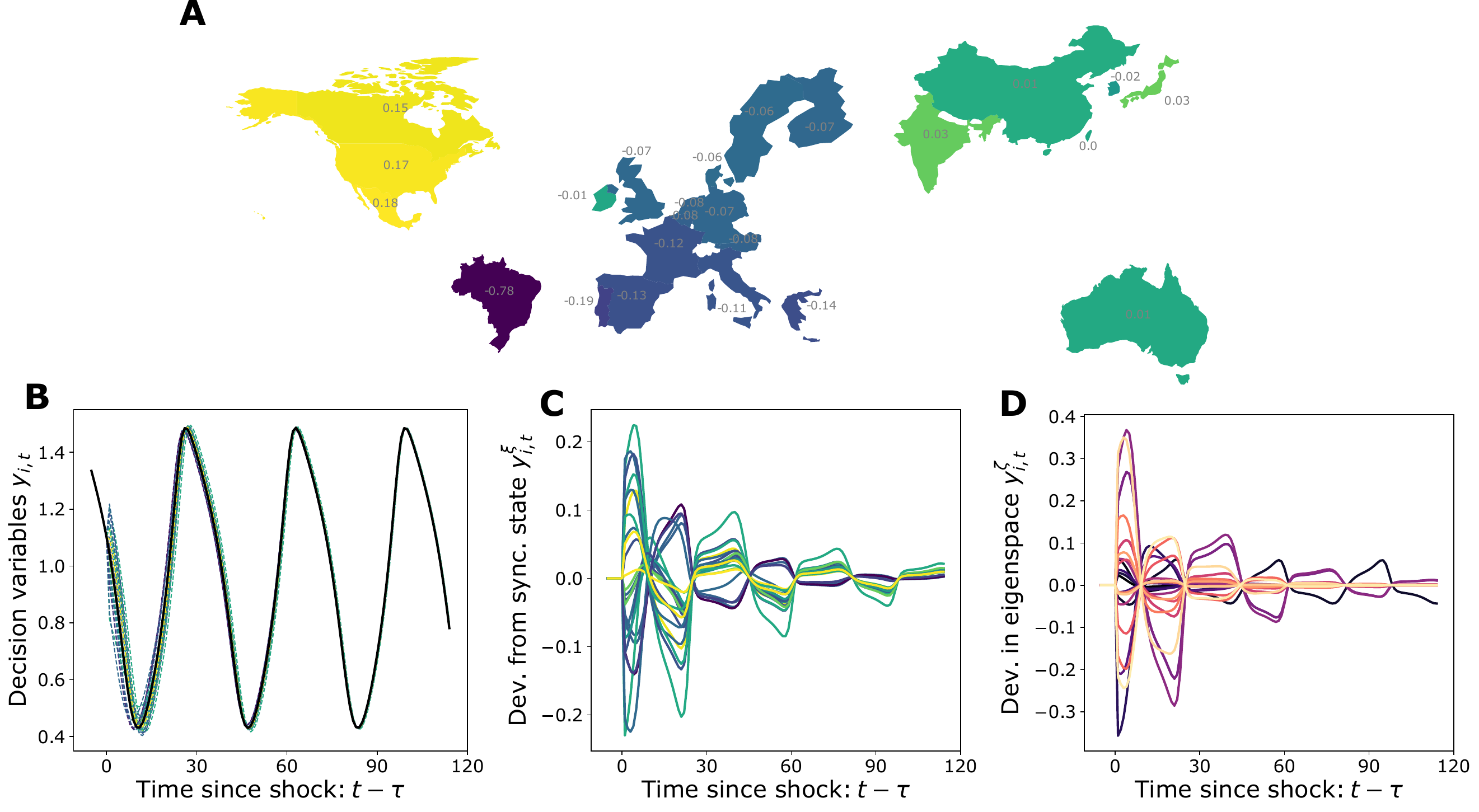}
\caption{\textbf{Master stability analysis applied to international synchronization.} A: Countries in our sample colored by their component of the Fiedler vector (also annotated). B: Dynamics of the decision variables $y_{i,t}$ in the synchronized state (solid black line), and in case of shocks (dashed lines representing individual countries, colored as in panel A). C: Deviations from the synchronized state after a shock is applied at time $\tau$. Each line represents a country and is colored like in panel A. D: Deviations projected in the eigenbasis. Each line represents an eigenmode; the darker the color, the smaller the corresponding eigenvalue is.}
\label{fig:network_plots}
\end{figure*}

To conclude this section, we now discuss a real-world example. We first explain how we build the  international trade network that specifies the coupling between countries, and then discuss the insights that can be gained by applying complete synchronization theory to these data.

\subsubsection{Trade network}
\label{sec:networkdata}

To build the network of interaction coefficients $\epsilon_{ij}$ described in Section \ref{sec:internationalmodel}, we merge the recently released Long-Run World Input-Output Database (LR-WIOD), covering the period 1965-2000 \citep{woltjer2021long}, with the established 2013 release of the WIOD, spanning the years 1995-2011 \citep{timmer2015illustrated}. We consider 24 countries (Figure \ref{fig:network_plots}A), comprising more than 80\% of World GDP, and a Rest of the World aggregate. Figure \ref{fig:network_visualization} in Appendix \ref{sec:supp_network_data} shows a visualization of the network in 2011.

We use international input-output tables instead of trade data because they make it easier to consistently estimate the share of internal demand vs. international demand. At the same time, we consider a long time span to test the robustness of our assumption of fixed interaction coefficients, which turns out to be a good approximation at business cycle frequencies (Appendix \ref{sec:supp_network_data}). The data also show that the share of internal demand is generally high (almost always above 0.7, often above 0.9), suggesting that the standard assumptions under which synchronization theory can be applied (weakly coupled self-sustaining oscillators) holds. 

\subsubsection{Master stability analysis}

Figure \ref{fig:network_plots}A shows the countries in our sample, colored by their component of the Fiedler vector of the matrix $\boldsymbol I_N-\boldsymbol \epsilon$, where $\boldsymbol I_N$ is the identity matrix of size $N$ and $\boldsymbol \epsilon$ is the matrix with components $\epsilon_{ij}$ (see Appendix \ref{sec:supp_real_networks}). The Fiedler vector is commonly used in graph partitioning because it is an approximate solution to the problem of dividing a graph into two components such that the sum of the links between the two components has minimal weight. When partitioning a graph using the Fiedler vector, one puts all nodes with positive components in one cluster and all nodes with negative components in the other cluster. Components that are close to zero in either the positive or negative direction are less clearly in one cluster than components that are far from zero.

Here, we see that Brazil's component is strongly negative, and no other country has so strongly negative values (presumably because Brazil trades a lot with other South American countries, which are included in our sample as ``Rest of the World'' and not shown here). Next, European countries have components ranging between -0.19 and -0.06 (with the exception of Ireland, which has -0.01), with Southern European countries having more negative values than Northern European countries. Furthermore, Asian countries have slightly positive components (with the exception of South Korea which is slightly negative). Finally, the components associated to North American countries range between 0.15 and 0.18. Thus, under graph partitioning one cluster would mostly comprise Brazil and Europe, and the other Asia and North America. 

To explore how the theory developed in Section \ref{sec:masterstabilityanalysismain} helps understand synchronization, we let the model reach the synchronized state, and then apply at a given time $\tau$ an i.i.d. shock to each country, drawn from a Gaussian distribution with mean zero and standard deviation 0.1. Such shocks are about 10\% of the amplitude of the endogenous limit cycles.

As we can see in Figure \ref{fig:network_plots}B, the dynamics of individual countries diverge from the synchronized state, but then after about 60 time steps they synchronize again. Figure \ref{fig:network_plots}C zooms into the deviations from the synchronized state, coloring each line by the same color of the corresponding country in panel A. Even if the shocks are independent across countries, we see that lines get roughly divided into two groups, along the lines of the graph partitioning mentioned above. 

Figure \ref{fig:network_plots}D shows the deviations from the synchronized state in the eigenbasis. We see that the black line, corresponding to the null eigenvalue, keeps fluctuating around zero, suggesting a small permanent shift of the shocked dynamics from the unshocked one. Moreover, dark lines corresponding to eigenmodes with small eigenvalues show damped fluctuations that take long to return to zero. These lines include the second eigenmode, implying that it takes longer to reach synchronization between the clusters defined by the Fiedler vector. Light-colored lines instead quickly return to zero, suggesting that synchronization within clusters happens fast.

In conclusion, the community structure discovered by our spectral analysis, combined with the theory of complete synchronization that we developed, makes it possible to understand analytically which countries are more likely to experience higher synchronization when each country follows an endogenous business cycle and is hit by idiosyncratic shocks.

\subsection{Comovement implications}
\label{sec:completesyncempirical}

Further to providing intuition into the mechanisms driving synchronization, the theory of complete synchronization helps explain how synchronization of endogenous business cycles could improve on exogenous business cycle models in terms of generating higher comovement between macroeconomic time series. 

Consider two countries following two deterministic identical limit cycles in isolation. When coupled, absent shocks, after some time their dynamics perfectly align in the synchronized state $\boldsymbol s_t=(x^s_t,y^s_t)$. Suppose now that at time $\tau$, the decision variables $y_{1,\tau}$ and $y_{2,\tau}$ of the countries are hit by idiosyncratic shocks $y_{1,\tau}^\xi$ and $y_{2,\tau}^\xi$ respectively. The comovement between the time series of the decision variables, calculated for the time steps $t$ immediately following the shocks, is given by the Pearson correlation coefficient
{\footnotesize
\begin{equation}
\label{eq:dynshock}
\begin{aligned}
\text{cor}\left(y^s_t+y_{1,t}^\xi,y^s_t+y_{2,t}^\xi\right)=\\
&\hspace{-100pt}\frac{\text{cov}\left(y^s_t,y^s_t\right)+\text{cov}\left(y^s_t,y_{2,t}^\xi\right)+\text{cov}\left(y_{1,t}^\xi,y^s_t\right)+\text{cov}\left(y_{1,t}^\xi,y_{2,t}^\xi\right)}{\text{std}\left(y^s_t+y_{1,t}^\xi\right)\cdot\text{std}\left(y^s_t+y_{2,t}^\xi\right)}.
  \end{aligned}
\end{equation}
}
In the above equation, cov denotes the covariance, std indicates the standard deviation, and we have used linearity of the covariance to decompose it in various terms. The term $\text{cov}\left(y^s_t,y^s_t\right)$ indicates the perfect synchonization of the unperturbed deterministic dynamics; the terms $\text{cov}\left(y^s_t,y_{2,t}^\xi\right)$ and $\text{cov}\left(y_{1,t}^\xi,y^s_t\right)$ indicate the comovement between the deterministic dynamics and the deviations due to the shocks; and the term $\text{cov}\left(y_{1,t}^\xi,y_{2,t}^\xi\right)$ indicates the correlation of the deviations (i.e., shock propagation). Compare this equation with the correlation between time series of two countries that are in the same stable steady state and are hit by shocks $(y_{1,t}^\xi, y_{2,t}^\xi)$:
\begin{equation}
\text{cor}\left(y_{1,t}^\xi,y_{2,t}^\xi\right)=\frac{\text{cov}\left(y_{1,t}^\xi,y_{2,t}^\xi\right)}{\text{std}\left(y_{1,t}^\xi\right)\cdot\text{std}\left(y_{2,t}^\xi\right)}.
\label{eq:steadyshock}
\end{equation}
If we assume that the variance of endogenous and exogenous fluctuations is the same, the correlation coefficient is only determined by the numerators of these expressions. Comparing \eqref{eq:dynshock} and \eqref{eq:steadyshock}, it is clear that the endogenous component of dynamics potentially increases correlation. In particular, when the variance of the shocks is small relative to the variance of $y^s_t$ (i.e., the typical amplitude of the endogenous cycles), the correlation coefficient tends to one when business cycles are purely endogenous.

\section{Empirical application}
\label{sec:empirical}

We now test quantitatively to which extent exogenous and endogenous business cycle models are able to match the empirical level of international comovement that can be found in real-world time series.

\subsection{Time series data}
\label{sec:tsdata}

To do so, we must measure empirical comovement in the first place. This is not a trivial task. Papers measuring business cycle comovement have been using all sorts of variables that measure economic activity and all sorts of filters to detrend the corresponding time series \citep{de2008will}. 

Here, we consider annual data on employment and value added to measure economic activity. We obtain these data from the Penn World Tables, version 10.0 \citep{feenstra2015next}. More specifically, we use the number of persons engaged as a measure of employment, and real GDP at constant 2017 national prices to measure value added. For the countries considered in this paper, data are available from 1950 to 2019. 

We cannot use these raw data to measure comovement, because they show clear common trends (confirmed by an augmented Dickey-Fuller test) that would bias upward the estimate of comovement at business cycle frequencies. To address this issue, we consider various modifications. First, we may divide employment or value added by total or working age population.\footnote{Total population is available in the Penn World Tables, while working age population is downloaded from the World Bank.} Second, we filter the resulting data so as to remove long-run fluctuations from the sample. To do so, we use two filters. One is the well-known Hodrick-Prescott filter, with both the standard parameter $\lambda=100$ that is used for annual data, and the value $\lambda=6.25$ suggested by \cite{ravn2002adjusting}. The other is the band-pass filter proposed by \cite{christiano2003band}, alternatively keeping fluctuations between 2-15 or 2-25 years.\footnote{We prefer the filter by \cite{christiano2003band} over the \cite{baxter1999measuring} filter because the latter loses some data points at the extremes of the series.} For all these options for the filters, we either consider the cyclical component or the ratio between the cyclical and trend components. In conclusion, from all these combinations we get 27 alternative detrending procedures for both employment and value added.

We find that the mean pairwise Pearson correlation among countries' employment is 0.23, with a standard deviation of 0.03 across detrending procedures. When considering value added, the correlation is 0.34, with a 0.05 standard deviation. These results are in line with estimates in the literature \citep{kose2006can}.

\subsection{Main result}
\label{sec:mainresult}

\begin{figure}[t]
\centering
\includegraphics[width=0.45\textwidth]{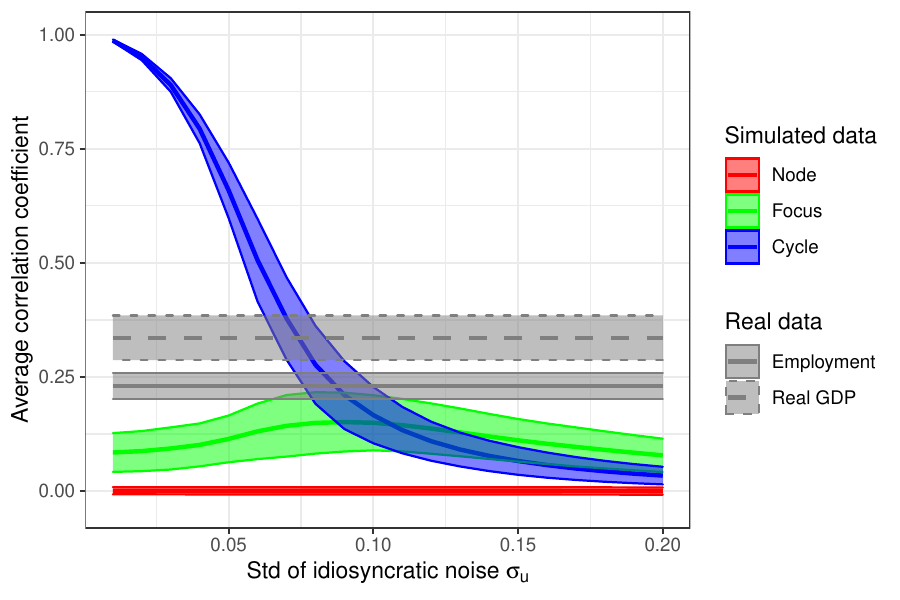}
\caption{Mean Pearson correlation across countries as a function of the standard deviation of idiosyncratic shocks, for three parameterizations of the model and for two empirical variables. For the simulations, each line represents the mean across 100 replications, and error bands represent one standard deviation. For the data, lines represent the mean across 18 detrending procedures, and error bands represent one standard deviation (Section \ref{sec:tsdata}). }
\label{fig:baseline_correlations_average}
\end{figure}

We now check whether endogenous business cycles are indeed necessary to reproduce the empirical level of comovement, or if instead shock propagation in an exogenous business cycle specification is sufficient. To do so, we simulate the model under three different parameterizations, and compare the comovement of the decision variables $y$ to empirical comovement as measured in the previous section. We perform this comparison at the aggregate, country and country-pair level.

We simulate the model under three baseline sets of parameters (many more parameter combinations are checked in Appendix \ref{sec:otherpars}). These correspond to the deterministic skeleton of the model producing limit cycles, or converging back to the steady state with damped fluctuations (\textit{focus}) or monotonically (\textit{node}). We choose the parameters in Appendix \ref{sec:numerical_example} as a baseline. Moreover, we parameterize the shock process as an AR(1) with persistence parameter $\rho=0.3$ and standard deviation $\sigma$ varying between 0.01 and 0.2. We simulate the model for 280 time steps after removing an initial transient. It makes sense to interpret time steps as quarters (Appendix \ref{sec:numerical_example}), so this leads to a 70-years period that is as long as the data we compare against (1950-2019).  To calibrate the interaction network, we consider year 1990, which is in the middle of the available data (we will also consider time-varying interaction coefficients as a robustness test).

Figure \ref{fig:baseline_correlations_average} shows our key result. When the model produces limit cycles, the mean correlation coefficient across countries ranges from 0.99 in the case of idiosyncratic shocks with standard deviation $\sigma_u = 0.01$ (or 1\% of the steady state value of the decision variables) to 0.03 when the standard deviation of idiosyncratic shocks is $\sigma_u = 0.20$ (20\% of the steady state). By contrast, when the model produces fluctuations that converge to a stable steady state, the average correlation coefficient is never larger than 0.15, for any value of the standard deviation of idiosyncratic shocks. More specifically, when the steady state is a focus, the average correlation varies between 0.08 and 0.15, whereas when the steady state is a node it is close to zero. As empirical comovement is 0.23 when considering employment and 0.34 when considering value added, only in the limit cycle case the model can reproduce this level of comovement. This is confirmed by a t-test for the difference in means, with high statistical significance for all values of $\sigma_u$ ($t$-statistic $<-11$). In conclusion, under the assumptions in this section the model can only match the empirical level of comovement with endogenous business cycles, for standard deviation of idiosyncratic shocks between 5 and 10\% of the amplitude of the limit cycles. This confirms the theoretical intuition in Section \ref{sec:completesyncempirical}.

\begin{figure*}[h!]
\centering
\includegraphics[width=1\textwidth]{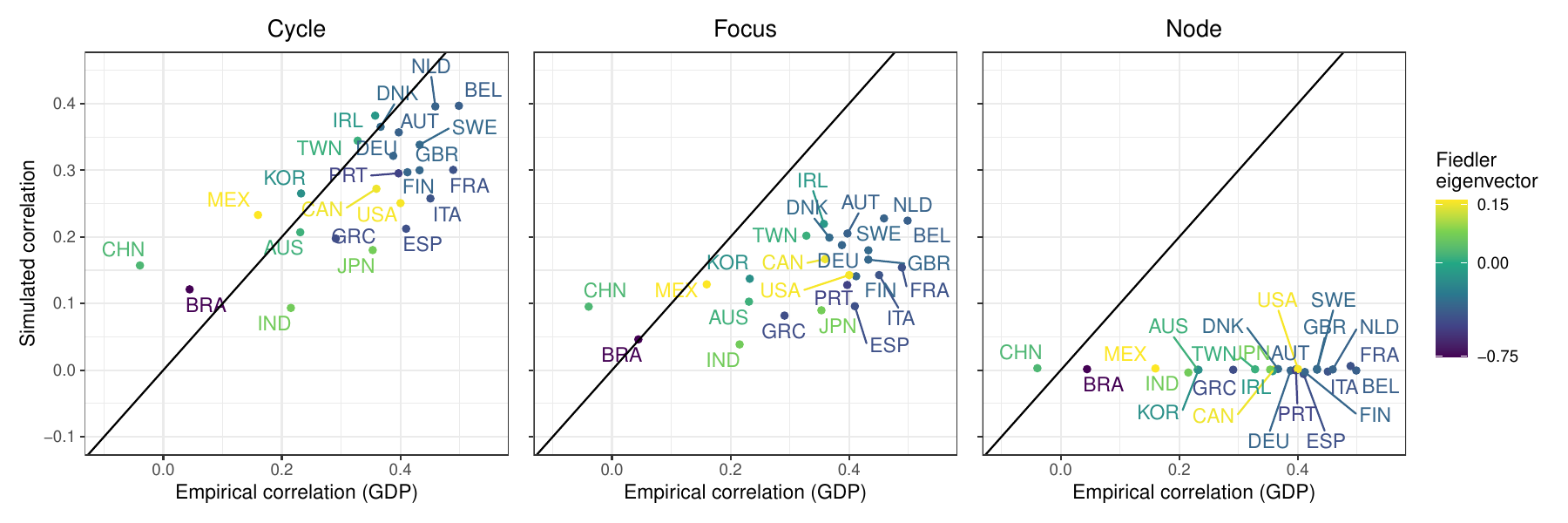}
\caption{Each point represents how much a given country is correlated to the other countries in the sample. The position on the horizontal axis indicates this country-specific correlation in the empirical data as measured from GDP, and is the same in the three panels. The vertical axis shows the same country-specific correlation in simulated data, as measured from the decision variables $y$, and is different in each panel depending on the model parameterization. All simulated results are averaged across 100 replications. The color code follows the position of the countries in the interaction network as exemplified by the Fiedler eigenvector, as in Figure \ref{fig:network_plots}. The identity line is plotted for reference. }
\label{fig:baseline_correlations_across_countries}
\end{figure*}

Next, Figure \ref{fig:baseline_correlations_across_countries} zooms in at the level of countries. Here, we select a standard deviation of idiosyncratic shocks of 0.08, under which the limit cycle model produces a correlation that is close to the data (Figure \ref{fig:baseline_correlations_average}). This figure conveys three results. First, for every country and not only for some, comovement is higher in the limit cycle case than in the focus and node regimes. Second, under limit cycles, empirical correlation coefficients are more similar to simulated ones (Pearson empirical-simulated coefficients: 0.69). This means that countries whose dynamics are more correlated to those of other countries in the data are also more correlated in the model. For instance, the mean correlation coefficient of the Netherlands is 0.46 in the data and 0.40 in the model, while the mean correlation coefficient of Brazil is 0.04 in the data and 0.12 in the model. This similarity between empirical and simulated correlation coefficients extends to the focus case, although it is slightly lower (Pearson 0.62). It is however completely absent when the steady state is a node (Pearson -0.15). The third result is that, in the limit cycle and focus cases, both empirical and simulated correlations are clearly linked to the position of the countries in the international trade network, as exemplified by the Fiedler eigenvector discussed in Section \ref{sec:real}. Countries that are more correlated to the other countries in the sample tend to have intermediate values for their eigenvector components, whereas countries with extreme values such as Brazil tend to have lower average correlations. 

\begin{figure*}[h!]
\centering
\includegraphics[width=1\textwidth]{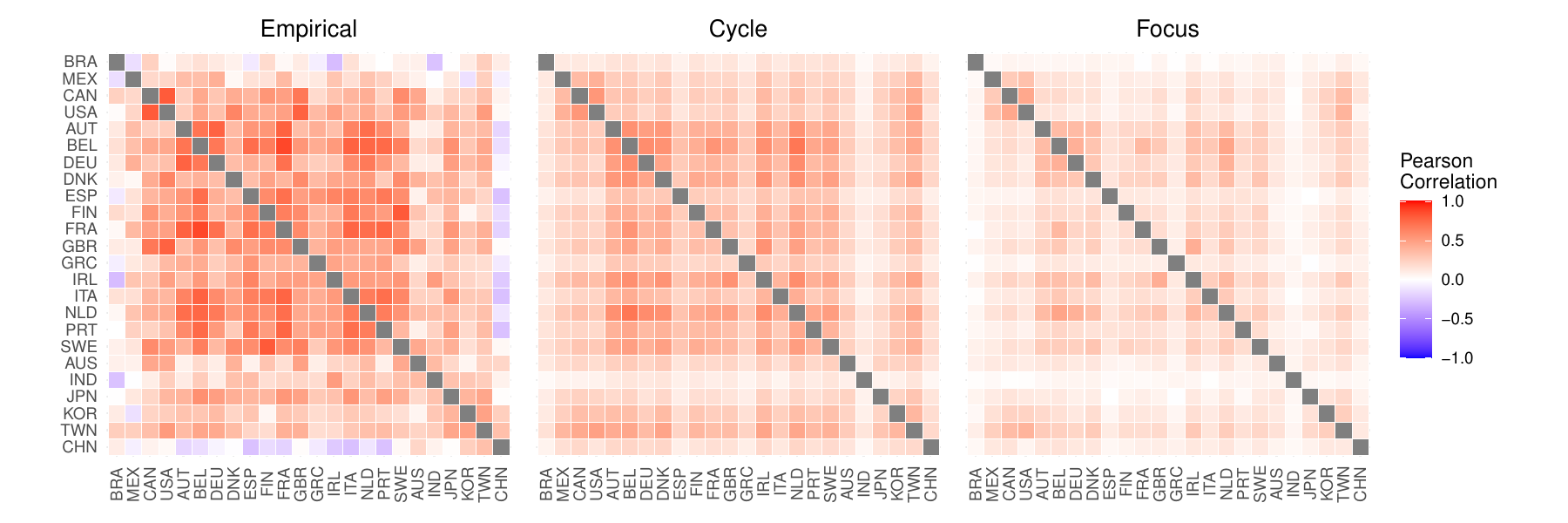}
\caption{Pairwise Pearson correlation coefficients between the dynamics in different countries, both in the model and in the data, averaged across 100 replications.}
\label{fig:baseline_pairwise_correlations}
\end{figure*}

These results are confirmed in Figure \ref{fig:baseline_pairwise_correlations}, which shows pairwise correlations. In the data, pairwise correlations are high within Europe, North America and Asia, but they are lower across continents. The simulations, both in the cycle and focus cases, generally reproduce these patterns, although they fail to yield the few negative correlations that can be seen in data. By contrast, when the model is parameterized as a node, pairwise correlations do not show a clear pattern (Figure \ref{fig:baseline_pairwise_supp} in Appendix \ref{sec:pairwise_supp}).

In conclusion, the results in this section support the main claim of the paper, namely that endogenous business cycles are needed to explain the level of international comovement that can be found in the data. They also show that the theory developed in Section \ref{sec:completesync} can be fruitfully employed to interpret synchronization.

\subsection{Robustness tests}
\label{sec:robustnessmain}

In Appendix \ref{sec:robustness} we check if the main result of the paper is robust to alternative specifications. We perform seven robustness tests. First, we investigate how higher persistence of the shock processes $\rho$ may influence the results. We find that, with high persistence, the model becomes more noisy for a given level of the standard deviation $\sigma$. So, in the limit cycle case, smaller values of  $\sigma$ are needed for the model to match the empirical level of comovement. Still, the focus and node cases cannot match empirical comovement for any value of $\rho$ and $\sigma$. The second test is to relax the assumption that the interaction coefficients $\epsilon_{ij}$ have to be fixed. In particular, we consider time-varying interaction coefficients $\epsilon_{ij,t}$, with $t=1965, \ldots, 2011$, representing varying shares of international trade over the data availability period. We find virtually identical results to the baseline. The third robustness test considers alternative parameters for the baseline specification of the interaction function $F$, which is a generic quartic function. It shows that in no case the model can match empirical comovement under focus or node parameterizations, whereas it can always do so with limit cycles, for an appropriate level of noise. The results are similar in the fourth robustness test, which considers an alternative logistic specification of the function $F$. The fifth robustness test considers other bifurcation parameters than the strength of strategic complementarities at the steady state. It still shows that node and focus parameterizations cannot produce a level of comovement that is as high as in the data, confirming that it is the type of dynamics and not the strength of strategic complementarities that substantially changes the level of comovement. The sixth robustness test shows that values of the $\alpha_1$ and $\alpha_2$ parameters that are heterogeneous across countries, although relatively close to the baseline, do not lead to almost any difference in the results. The seventh, and last, robustness test considers common shocks. We find that for sufficiently strong common shocks the focus parameterization does lead to sufficiently high comovement, but the node parameterization always fails to do so.

\section{Conclusion}
\label{sec:conclusion}

This paper contributes to the debate on whether business cycles are exogenous or endogenous. On an empirical level, it has not been possible to find conclusive evidence for limit cycles or chaos in economic time series, but it has not been possible to rule them out either. One contribution of this paper is to propose an indirect empirical test, as comovement of economic time series has different origins when business cycles are exogenous vs. endogenous, originating from shock propagation in the first case and from synchronization in the second case. Our main empirical result is that a limit cycle model, buffeted by relatively small idiosyncratic shocks, comes closest to explaining the level of comovement that can be found in macroeconomic time series. 

This result comes with several caveats. First, we consider a reduced-form model that interprets the linkages between countries as international trade, rather than a full-fledged structural model. This prevents us from estimating deep parameters and shock processes. Second, the unit of our model is a country, but an interesting extension would be to consider sector-country pairs, as it would allow to consider the impact of sector-specific shocks. Third, we mostly neglect common shocks, as it would be trivial to obtain a level of comovement that matches the data if the standard deviation of the common shocks is sufficiently large; in our view, to properly evaluate the role of common shocks, they would need to be estimated with a structural model. Despite these limitations, we expect our main result to hold in more complicated models: when deterministic underlying dynamics synchronize, comovement is bound to be larger.

On a theoretical level, we extend complete synchronization theory to deal with an economic problem. The idea that synchronization can help explain comovement has been considered \citep{krugman1996self}, but never tested empirically. Moreover, almost everyone thinking about synchronization considers phase synchronization. Using complete synchronization, we do not require that the underlying time series have a clear periodicity, and we investigate the relative role of exogenous shocks and endogenous non-linear dynamics in more detail. Beyond macroeconomics, we hope that the tools that we offer in this paper can find wider applicability, as any disaggregate economic and financial model that can be described by some form of non-linear dynamics can be studied with the tools of synchronization theory.

%In a popular science book, \cite{krugman1996self} wrote: ``One of the luxuries of a format like this one is that I can include the kind of loose speculations that I could never write in a journal and that I can explain, as I am doing now, that I do not necessarily believe in the theory I am advancing. So here is a crazy idea about the global business cycle: it is an example of ``phase locking.''[...] Like the two back-to-back clocks that started ticking in unison, the two economies would not need to be very strongly linked to develop a synchronized cycle; a modest linkage would do as long as they were predisposed to have cycles in any case and had fairly similar natural periods.'' 

%In this paper, we seriously explore this hypothesis, showing that it is highly promising to explain comovement of business cycles across countries. Differently from Krugman, however, we do not put much emphasis on phase locking and periodicity, which are problematic in the analysis of economic time series, but we rather develop a theory that explains synchronization of generic non-linear dynamics of nodes in an interaction network that are hit by exogenous shocks.

\newpage
\appendix

\section{Analysis of uncoupled dynamics}
\label{sec:uncoupled_dyn_analysis}

\subsection{Homogeneous dynamics}
\label{sec:uncoupled_dyn}

Here we mathematically describe how the general framework for endogenous fluctuations laid out in Section \ref{sec:abstractmodel}  may generate cyclical dynamics (see \cite{beaudry2015reviving} for more details). For simplicity, we focus on a single country $i$ whose dynamics is purely driven by internal demand, namely $\epsilon_{ii}=1$, and $\epsilon_{ij}=0$, for all $j\neq i$ (autarchic case). This can be interpreted as the country being composed by a large number of agents with a homogeneous interaction network and we consider the symmetric solution in which all agents behave alike, as in \cite{beaudry2015reviving}. For ease of notation, in this section we drop the subscript $i$.

We first look for the steady state $(\hat{x},\hat{y})$. Following the characterization of $x$ and $y$ as oscillation variables, as discussed in Section \ref{sec:internationalmodel}, for convenience we constrain parameters so that $\hat{y}=1$ is the unique steady state. This means that the steady state value of the accumulation variable is $\hat{x}=1/\delta$ and the following relation must hold: $\alpha_0=1-\alpha_1/\delta-\alpha_2-F(1)$. In the following, we always select $\alpha_0$ so that this condition is satisfied. For the steady state to be unique, the slope of the line $-\alpha_0+y_t(1-\alpha_1/\delta-\alpha_2)$ must be larger than the derivative of $F$ at $\hat{y}=1$, namely $1-\alpha_1/\delta-\alpha_2>F'(1)$, suggesting that strategic complementarities at the steady state should not be too large. We always choose parameters so that the steady state is unique.

\begin{figure}
\centering
\includegraphics[width=0.8\textwidth]{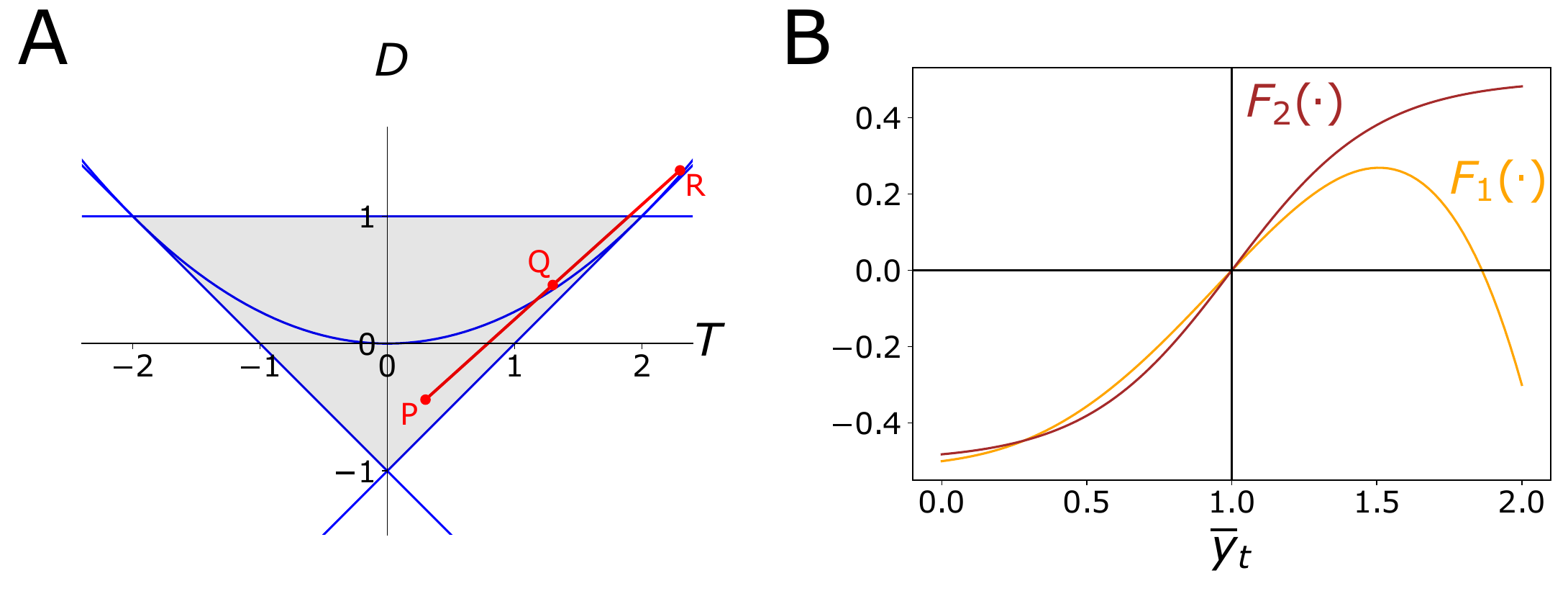}
\caption{Stability of dynamics for homogeneous agents. (A) Diagram for stability of 2-dimensional maps. The blue lines correspond to $D=1$, $D=T-1$, $D=-T-1$ and $D=(T)^2/4$. The red line corresponds to varying $F'(1)$ with the other parameters fixed. (B) Specifications $F_1(\cdot)$ and $F_2(\cdot)$ for $\overline{y}_t\in[0,2]$ (see Appendix \ref{sec:numerical_example}).}
\label{fig:stability_model}
\end{figure}

The Jacobian of this system is 
\begin{equation}
\mathcal J = \begin{pmatrix}
  1 - \delta & 1  \\
  \alpha_1 & \alpha_2+F'(1)
 \end{pmatrix}.
\label{eq:jacobian_end_bc}
\end{equation}
The stability of the steady state is completely characterized in terms of the trace $T=1-\delta+\alpha_2+F'(1)$ and determinant $D=(1-\delta)\left(\alpha_2+F'(1)\right)-\alpha_1$ of the Jacobian. Following the standard conditions for 2-dimensional maps, stability obtains if $D<1$, $D>T-1$ and $D>-T-1$ (the grey region in Figure \ref{fig:stability_model}A, representing the usual diagram for the stability of 2-dimensional maps). If $D>(T)^2/4$, the eigenvalues of the Jacobian are complex, and so the system admits either damped oscillations or sustained cycles. If the line $D=1$ is crossed above $D=(T)^2/4$, the system undergoes a Hopf (Neimark-Sacker) bifurcation. \cite{beaudry2015reviving} show that the Hopf bifurcation is supercritical, i.e. the resulting limit cycle is attractive.

To build some intuition into the transition between stability and instability, we fix the parameters $\alpha_1$, $\alpha_2$, and $\delta$, and vary the values of $F'(1)$ so that it varies in the interval $F'(1)\in[-1,1]$. In other words, we consider the level of complementarity or substitutability at the steady state as a bifurcation parameter. Point P in Figure \ref{fig:stability_model}A corresponds to $F'(1)=-1$. This corresponds to strategic substitutability, and indeed under this parameterization the steady state is a stable node, meaning that eigenvalues are real and of magnitude smaller than one. When the steady state is a node, dynamics converge without oscillations. Next, point Q corresponds to $F'(1)=0$, i.e. to a transition between strategic complementarity and substitutability. Under these parameters, the system has a stable focus, with complex eigenvalues of magnitude smaller than one. In this setting, dynamics converge to the steady state producing damped oscillations.  As the level of complementarities increases, the steady state loses stability through a Hopf bifurcation, up to point R, corresponding to $F'(1)=1$.

The analysis so far has focused on local stability of the steady state. To build some intuition into global dynamics, in Figure \ref{fig:stability_model}B we plot the function $F(\cdot)$ for two specifications $F_1(\cdot)$ and $F_2(\cdot)$ detailed in the next section. Under both parameterizations, the function $F$ increases at the steady state, $\overline{y}_t=1$, but flattens out or even decreases when $\overline{y}_t$ becomes large, which leads to smaller complementarities or even to strategic substitutability and prevents explosive dynamics.

\subsection{A numerical example}
\label{sec:numerical_example}

We now show representative time series for the various cases discussed above: stable steady state of type node and focus, and limit cycle. 

To perform numerical simulations, we must specify the function $F$. As our main specification, we consider a generic quartic function, as it is a very flexible functional form that can have weak or strong complementarity, or even substitutability, at the steady state. Likewise, it may flatten out or even decrease far from the steady state. Thus, our first specification $F_1$, in the most general terms, is given by
\begin{equation}
F_1(y)=\beta_0 + \beta_1 y + \beta_2 y^2 + \beta_3 y^3 + \beta_4 y^4.
\label{eq:F1}
\end{equation}
Depending on the values of the $\beta$ parameters, the steady state $\hat{y}=1$ can be a node, a focus, or may be unstable. (In all cases, we select the $\beta$ parameters so that, conveniently, it is $F_1(1)=0$, as in \cite{beaudry2015reviving}). 

For robustness, we also consider a logistic function as an alternative specification: 
\begin{equation}
F_2(y)=\frac{1}{1+e^{-\beta (y-1)}}-\frac{1}{2}.
\label{eq:F2}
\end{equation}
In the above equation, the term $(y-1)$ ensures that the steady state is $\hat{y}=1$, and the shift $1/2$ makes sure that $F_2(1)=0$, as in Eq. \ref{eq:F1}. The larger $\beta$, the stronger the complementarities at the steady state.

Moreover, we select the $\alpha_1$, $\alpha_2$ and $\delta$ parameters for the limit cycle specification so that every cycle lasts about 36 time steps. If we interpret time steps as quarters, this is in line with the empirical evidence that the US economy tends to undergo a business cycle every about 9 years \citep{beaudry2018putting}. Experimenting with larger values of $\alpha_1$ (higher dislike for accumulation), we found out that the model could exhibit chaotic dynamics.

In conclusion, we consider the following baseline parameterizations:
\begin{itemize}
\item Node: $\alpha_1=-0.04$, $\alpha_2=0.4$, $\delta=0.1$, $\beta_0=-0.19$, $\beta_1=-0.11$, $\beta_2=0.4$, $\beta_3=0.2$, $\beta_4=-0.3$.
\item Focus: $\alpha_1=-0.04$, $\alpha_2=0.4$, $\delta=0.1$, $\beta_0=-0.2$, $\beta_1=-0.1$, $\beta_2=0.1$, $\beta_3=0.3$, $\beta_4=-0.1$.
\item Cycle: $\alpha_1=-0.04$, $\alpha_2=0.4$, $\delta=0.1$, $\beta_0=-0.5$, $\beta_1=0.1$, $\beta_2=0.2$, $\beta_3=0.5$, $\beta_4=-0.3$.
\item Chaos: $\alpha_1=-0.35$, $\alpha_2=0.4$, $\delta=0.1$, $\beta_0=-0.5$, $\beta_1=0.1$, $\beta_2=0.2$, $\beta_3=0.5$, $\beta_4=-0.3$.
\end{itemize}

\begin{figure}[h!]
\centering
\includegraphics[width=0.8\textwidth]{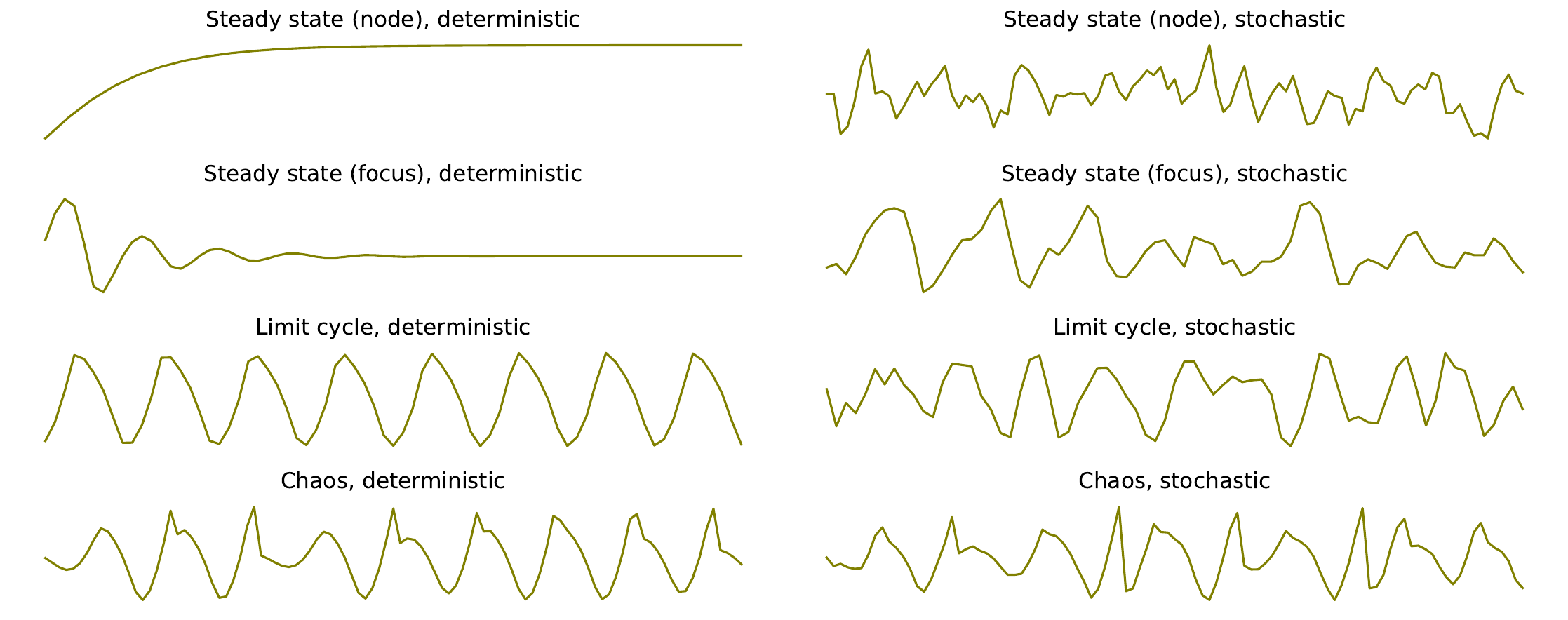}
\caption{Qualitatively different types of dynamics (node, focus, limit cycle and chaos), depending on the strength of strategic complementarities and other parameters, for both the deterministic and stochastic cases.}
\label{fig:types_business_cycles}
\end{figure}

We show the time series for each case in Figure \ref{fig:types_business_cycles}, showing both the corresponding deterministic and stochastic dynamics. (In all cases, the stochastic terms $u_{i,t}$ follow an AR(1) process with autocorrelation $\rho=0.3$ and standard deviation $\sigma=0.1$.) When the steady state is a node, deterministic dynamics converge to the steady state without oscillations, and the corresponding stochastic dynamics are quite irregular. When the steady state is a focus, convergence happens under damped oscillations, and the corresponding stochastic dynamics show more structure. Under limit cycles, deterministic dynamics  are perfectly periodic, while stochastic dynamics retain the shape of the cycle with clearer periodicity than under a focus. Finally, under chaotic dynamics, even in the deterministic case there is no perfect periodicity, and it is even more so in the stochastic case. 

In conclusion, the dynamical system in Eq. \eqref{eq:dynsyst} of the main paper exhibits a variety of dynamics, some that converge towards stable steady states, in line with the usual narrative of business cycles driven by exogenous shocks, and others that instead follow limit cycles or chaos, resulting in endogenous business cycles. When adding noise, all dynamics could qualitatively resemble business cycle fluctuations, which are clearly not perfectly periodic but do show some structure.

\section{Master stability analysis}
\label{sec:masterstabilityanalysis}

In this section, we provide the technical details to the master stability analysis discussed in the main text.

\subsection{Master stability functions}
\label{sec:completesynctheory}

In the following, we first derive our key equations, and then provide some examples that clarify the working of our approach. To be as general as possible, we mostly focus on the abstract formulation of the model (Section \ref{sec:abstractmodel}). 

Let $\boldsymbol s_t=(x^s_t,y^s_t)$ denote the synchronized state in which all agents behave alike. Let the dynamics of individual agents be given by $x_{i,t}=x^s_t+x^\xi_{i,t}$ and $y_{i,t}=y^s_t+y^\xi_{i,t}$, where $x^\xi_{i,t}$ and $y^\xi_{i,t}$ denote a small deviation from the synchronized state, for example due to idiosyncratic shocks hitting agent $i$. Let further $\boldsymbol \xi_t = \left(x^\xi_{1,t}, y^{\xi}_{1,t}, x^{\xi}_{2,t}, y^{\xi}_{2,t}, \ldots x^{\xi}_{N,t}, y^{\xi}_{N,t} \right)$ be the 2$N$-dimensional vector of deviations. 

To compute the evolution of this vector, we take advantage of the small deviations to Taylor-expand the non-linear terms in the dynamical equations. We illustrate this with a specific example of a star network with three nodes. This network is composed by a central node, which we denote as node 1, that is linked to two leaf nodes, which we denote as nodes 2 and 3. The leaf nodes are only connected to the central node. We assume that the dependence on other nodes is $\epsilon_{ij}=\epsilon/k_i$, where $k_i$ is the degree of node $i$, while the dependency on oneself is $\epsilon_{ii}=1-\epsilon$. This defines a weighted interaction network with adjacency matrix
\begin{equation}
\begin{pmatrix}
1-\epsilon & \epsilon/2 & \epsilon/2 \\
\epsilon & 1-\epsilon & 0 \\
0 & \epsilon & 1-\epsilon
\end{pmatrix}.
\end{equation}  
Replacing $x_{i,t}=x^s_t+x^\xi_{i,t}$ and $y_{i,t}=y^s_t+y^\xi_{i,t}$ in Eq. \eqref{eq:dynsyst} with the interaction network above, we get the following system of equations for the evolution of $\boldsymbol \xi_t$:

\begin{equation}
\begin{cases} 
x^{\xi}_{1,t+1} = (1-\delta)x^{\xi}_{1,t} + y^{\xi}_{1,t}, \\ 
y^{\xi}_{1,t+1} = \alpha_1 x^{\xi}_{1,t} + \left(\alpha_2 + F'(y^s_t) \right) y^{\xi}_{1,t}-\epsilon F'(y^s_t) \left( y^{\xi}_{1,t}-\frac{1}{2}y^{\xi}_{2,t} -\frac{1}{2}y^{\xi}_{3,t} \right),\\
x^{\xi}_{2,t+1} = (1-\delta)x^{\xi}_{2,t} + y^{\xi}_{2,t}, \\ 
y^{\xi}_{2,t+1} = \alpha_1 x^{\xi}_{2,t} + \left(\alpha_2 + F'(y^s_t) \right) y^{\xi}_{2,t}-\epsilon F'(y^s_t) \left( y^{\xi}_{2,t}-y^{\xi}_{1,t} \right) ,\\
x^{\xi}_{3,t+1} = (1-\delta)x^{\xi}_{3,t} + y^{\xi}_{3,t}, \\ 
y^{\xi}_{3,t+1} = \alpha_1 x^{\xi}_{3,t} + \left(\alpha_2 + F'(y^s_t) \right) y^{\xi}_{3,t}-\epsilon F'(y^s_t) \left( y^{\xi}_{3,t}-y^{\xi}_{1,t} \right),
\end{cases}
\label{eq:master_stab_example1}
\end{equation}
where we have used the fact that the deviations $\boldsymbol \xi_t$ are small to Taylor-expand the function $F$ to first order. This formulation suggests that the evolution of deviations of any single agent is given by the Jacobian of the dynamics corresponding to that agent, plus a term of interaction with other agents. Extrapolating from this example, it is possible to see that the evolution of the whole vector $\boldsymbol \xi_t$ is generically given by
\begin{equation}
\boldsymbol \xi_{t+1}=\left(\boldsymbol I_N \otimes \boldsymbol J(\boldsymbol s_t) - \epsilon F'(\boldsymbol s_t) \boldsymbol K \boldsymbol L \boldsymbol K \otimes \boldsymbol H \right) \boldsymbol \xi_t,
\label{eq:masterstab}
\end{equation}
where:
\begin{itemize}
\item $\boldsymbol I_N$ is the $N$-dimensional identity matrix.
\item $\boldsymbol J(\boldsymbol s_t)$ is the 2-dimensional Jacobian of the dynamical system.
\item $\otimes$ denotes the Kronecker product, i.e. a 2$N$-dimensional matrix.
\item $F'(\boldsymbol s_t)$ is the first derivative of $F$, due to the fact that we are considering a first-order approximation around the synchronized state.
\item $\boldsymbol K$ is an $N$-dimensional square matrix with $1/\sqrt{k_i}$ on the main diagonal and zero everywhere else.
\item $\boldsymbol L$ is the Laplacian of the network. This is a key mathematical property of the network that is widely used in many applications. $L_{ii}=k_i$ and $L_{ij}=-1$ if $i$ and $j$ are connected. $\boldsymbol K \boldsymbol L \boldsymbol K$ is known as normalized Laplacian, and has similar properties to the Laplacian.\footnote{\label{foot:fiedler}
Here we summarize a few properties of $\boldsymbol K \boldsymbol L \boldsymbol K$. Because the rows sum to zero, one eigenvalue is zero. Moreover, it is well known that the other eigenvalues are positive and bounded between 0 and 2. The multiplicity of the 0 eigenvalue reflects the number of disconnected clusters in the network. We sort the eigenvalues in increasing order, so that $\lambda_1=0$, $\lambda_2>0$ if the network is connected, $\lambda_3\geq \lambda_2$, ..., $\lambda_N\geq\lambda_{N-1}$, $\lambda_N\leq 2$. The eigenvalue $\lambda_2$ is knows as \textit{algebraic connectivity} and the corresponding eigenvector as \textit{Fiedler vector}. The smaller $\lambda_2$, the more the network has a modular structure, in which two or more clusters of nodes have strong internal connectivity and weak external connectivity. In the case of two clusters, the components of the Fiedler vector are positive for nodes in a cluster and negative for nodes in the other cluster. Fiedler vectors are commonly used for graph partitioning.
}
\item $\boldsymbol H$ is the 2-dimensional square matrix of connectivity in the dynamical system. Since here the only connectivity is through $y$, it is $H_{22}=1$ and zero everywhere else.
\end{itemize}

The problem with Eq. \eqref{eq:masterstab} is that the evolution of each component $y_{i,t}^{\xi}$ depends on all other agents $j$ to which $i$ is connected. The key trick of the master stability approach is to diagonalize $\boldsymbol K \boldsymbol L \boldsymbol K$ so as to decompose the deviations $\boldsymbol \xi$ into orthogonal, uncoupled, components. Following the terminology in the literature, we call these components \textit{eigenmodes}. Let 
\begin{equation}
\boldsymbol \xi_t = \left( \boldsymbol Q \otimes \boldsymbol I_2 \right) \boldsymbol \zeta_t,
\label{eq:eigen}
\end{equation} 
where $\boldsymbol Q$ is the matrix of eigenvectors of $\boldsymbol K \boldsymbol L \boldsymbol K$. Here, $\boldsymbol \zeta_t$ can be interpreted as a projection of $\boldsymbol \xi_t$ in the eigenspace. Replacing Eq. \eqref{eq:eigen} in Eq. \eqref{eq:masterstab}, we see that $\boldsymbol \zeta_t$ evolves according to 
\begin{equation}
\boldsymbol \zeta_{t+1}=\left(\boldsymbol I_N \otimes \boldsymbol J(\boldsymbol s_t) - \epsilon F'(\boldsymbol s_t) \boldsymbol \Lambda \otimes \boldsymbol H \right) \boldsymbol \zeta_t,
\end{equation}
where $\boldsymbol \Lambda$ is the matrix with the eigenvalues of $\boldsymbol K \boldsymbol L \boldsymbol K$ on the main diagonal and zero everywhere else. Each of the $N$ eigenmodes of $\boldsymbol \zeta_{t}$ can then be written as
\begin{equation}
\boldsymbol \zeta_{i,t+1}=\left(\boldsymbol J(\boldsymbol s_t) -  F'(\boldsymbol s_t) \epsilon \lambda_i \boldsymbol H \right) \boldsymbol \zeta_{i,t}.
\label{eq:lyap}
\end{equation}
In this basis, the evolution of each eigenmode $i$ only depends upon itself. We stress that $i$ now is an eigenvector (corresponding to the eigenvalue $\lambda_i$ of the normalized Laplacian of the network) and \textit{not an agent}. As will be clarified in the examples below, the evolution of each eigenmode corresponds to higher-order properties of the network. Of course one can retrieve the dynamics of the agents by applying the transformation $\boldsymbol \xi_t = \left( \boldsymbol Q \otimes \boldsymbol I_2 \right) \boldsymbol \zeta_t$. 

The eigenmode $i=1$ corresponds to the eigenvalue $\lambda_i=0$, and describes dynamics parallel to the synchronization manifold (i.e., ``in the same direction as the dynamics''). It captures the phase shift due to a shock hitting the system at a given time $t$, as will be clarified below. The eigenmodes $i>1$ correspond to dynamics orthogonal to the synchronization manifold. If these dynamics always converge to zero after the shock, the synchronized state is stable. 

Whether the orthogonal dynamics converge to zero is not obvious from Eq. \eqref{eq:lyap}. To know if they do, one must numerically compute the Lyapunov exponents of the system. Letting $K=\epsilon \lambda_i$ denote an effective coupling for eigenmode $i$, one calculates Lyapunov exponents $\mu$ in Eq. \eqref{eq:lyap} for all values of $K$ that are typically obtained. (In our case, because $\epsilon\in[0,1]$ and $\lambda_i\in[0,2]$, it is of interest to only consider the interval $K\in[0,2]$.) When Lyapunov exponents are negative, the corresponding eigenmodes die out over time. 

\begin{figure}[h!]
\centering
\includegraphics[width=0.8\textwidth]{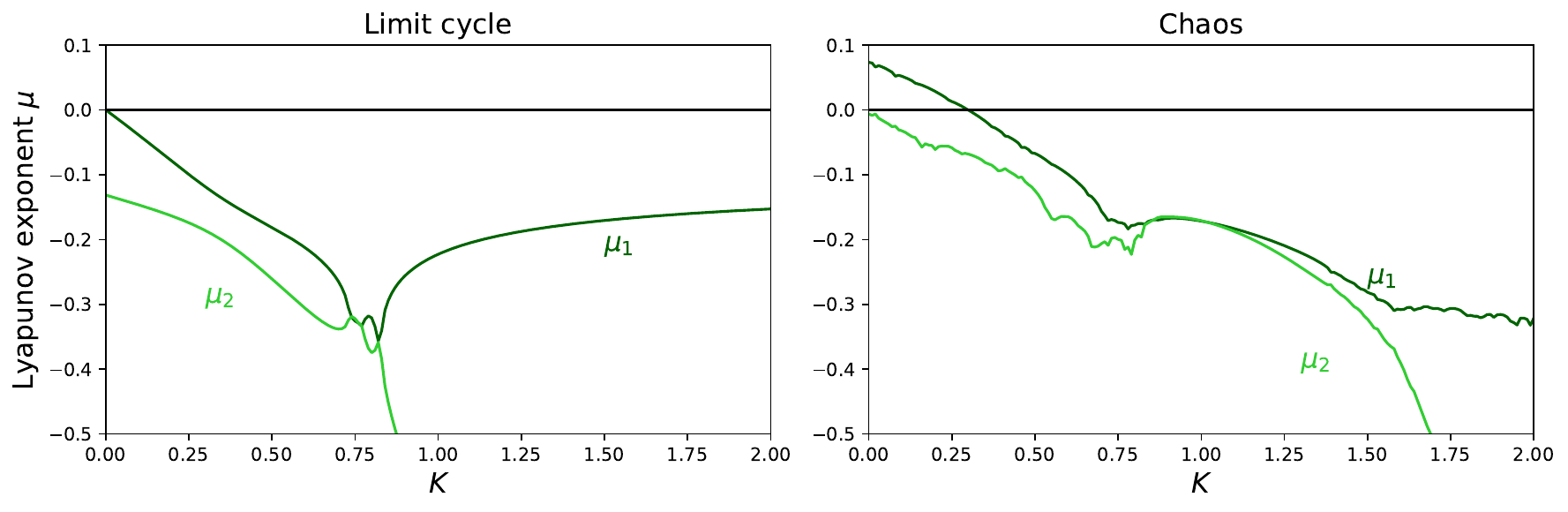}
\caption{Lyapunov exponents $\mu_1$ and $\mu_2$ as a function of the effective coupling $K$, both for a parameter combination that leads to a limit cycle and for a parameter combination that leads to chaotic dynamics. }
\label{fig:master_stability_function}
\end{figure}

The Lyapunov exponents are shown in Figure \ref{fig:master_stability_function}, for the baseline parameterizations of the model that lead to limit cycles and chaotic dynamics (Appendix \ref{sec:numerical_example}). There are two Lyapunov exponents because the dynamical system is 2-dimensional. Under limit cycles, both Lyapunov exponents $\mu_1$ and $\mu_2$ are always negative, except for $K=0$ in which case $\mu_1=0$. This means that all eigenmodes $i>1$ die out, making the synchronized state stable. In other words, if countries followed deterministic limit cycles and were only occasionally hit by idiosyncratic shocks, their business cycles would perfectly synchronize some time after each shock. Under chaotic dynamics, there is a region under which $\mu_1>0$, meaning that, even without exogenous shocks, under some conditions countries' business cycles would not synchronize. 

\subsection{Examples}
\label{sec:examples}

To clarify the formalism introduced in the previous section, we now consider two examples with the abstract framework for business cycles and a simple interaction network. For simplicity, in all cases we focus on limit cycles rather than chaotic dynamics, but the following analysis is analogous in the chaotic case. Moreover, we study the effect of single idiosyncratic shocks hitting the agents in the network at a certain time $\tau$, and study its propagation. 

\subsubsection{Two connected agents}
\label{sec:two}

We start from the simplest possible system to study the interplay between dynamics and shocks in most detail, disregarding the network structure. We consider two connected agents, agent 1 and agent 2, giving weight $1-\epsilon$ to their internal dynamics and $\epsilon$ to the dynamics of the other agent. We simulate the two agents until they reach the synchronized state (which they always do given that we consider limit cycle dynamics), and then hit agent 1 with a positive shock. We first show the numerical simulations, and then make sense of the results using the formalism developed in Section \ref{sec:completesynctheory}.

\begin{figure}[h!]
\centering
\includegraphics[width=1\textwidth]{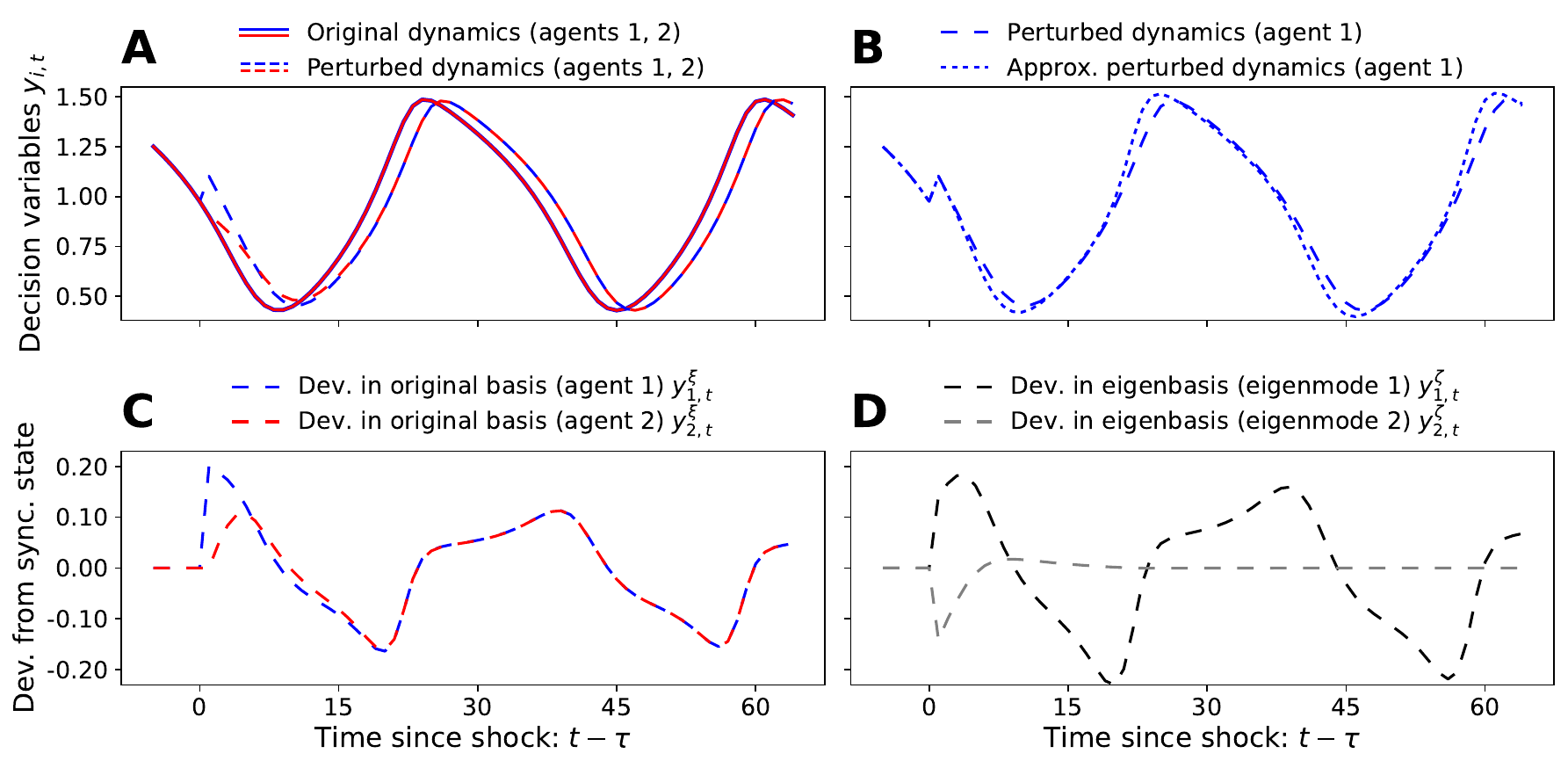}
\caption{Example dynamics: two connected agents are synchronized in a limit cycle, and at time $\tau$ agent 1 gets hit by a positive shock. We show the resulting dynamics at time $t>\tau$. A: Original (unshocked) and perturbed dynamics of agent 1 (blue) and agent 2 (red). B: Perturbed dynamics of agent 1 (as in panel A) compared with the linearized approximation $y_t^s+y_{i,t}^\xi$. C: Deviations from the synchronized state in the original basis, for both agent 1 ($y_{1,t}^\xi$) and agent 2 ($y_{2,t}^\xi$). D: Deviations from the synchronized state in the eigenbasis, for both eigenmode 1 ($y_{1,t}^\zeta$) and eigenmode 2 ($y_{2,t}^\zeta$). }
\label{fig:example_two_nodes}
\end{figure}

Figure \ref{fig:example_two_nodes} shows the dynamics in the 60 time steps following the positive shock to agent 1. Panel A compares the original (unshocked) to the perturbed dynamics, for both agents 1 and 2. It is clear that in the perturbed dynamics the two agents quickly reach synchronization again, but their business cycles permanently lead those of the original dynamics by a few time steps. Next, panel B shows how good the approximation of Eq. \eqref{eq:masterstab} is. In this panel, the perturbed dynamics of agent 1 (same as previously shown in panel A) is compared to the linearized approximation $y_t^s+y_{i,t}^\xi$. We see that that the exact and approximate dynamics are not identical, especially at the peaks and troughs, but overall the approximation is pretty faithful.

Having established this, we focus on the dynamics of the shocks in the original basis and in the eigenbasis. In the original basis (panel C), the terms $y_{1,t}^\xi$ and $y_{2,t}^\xi$ represent the deviations of agent 1 and agent 2 from the synchronized state. We see that, after agent 1 is hit by a positive shock at $\tau$ and deviates positively from the synchronized state, also agent 2 deviates positively from the synchronized state after a few time steps. Then, after a transient, the deviations from the synchronized state become identical across the two agents and start following a limit cycle. In the eigenbasis (panel D), the deviations $y_{1,t}^\zeta$ and $y_{2,t}^\zeta$ correspond to the first and second eigenmode. While the first eigenmode starts following a limit cycle similar to the one in panel C, the second eigenmode converges to zero after a negative fluctuation which then becomes slightly positive.

To make sense of these results analytically, we now consider the master stability formalism. For this basic network, the eigenvalues of $\boldsymbol K \boldsymbol L \boldsymbol K$ are $\lambda_1=0$ and $\lambda_2=2$. Looking at Figure \ref{fig:master_stability_function} (left panel), the first eigenmode, corresponding to $\lambda_1=0$ and so $K=0$, has largest Lyapunov exponent $\mu_1=0$. Thus, from Eq. \eqref{eq:lyap}, it is $\boldsymbol \zeta_{1,t+1}=\boldsymbol J(\boldsymbol s_t) \boldsymbol \zeta_{1,t}$: the dynamics of the first eigenmode are always described by the Jacobian of the synchronized state (black line in Figure \ref{fig:example_two_nodes}D). By contrast, the second eigenmode, corresponding to $\lambda_2=2$ and so $K=0.6$,\footnote{In this example, we take $\epsilon=0.3$.}, has largest Lyapunov exponent $\mu_1\approx -0.2$. This means that it exponentially converges to zero, approximately by 20\% every time step (grey line in Figure \ref{fig:example_two_nodes}D).

The eigenvectors corresponding to the two eigenmodes are $\boldsymbol v_1=(1,1)/\sqrt{2} $ and $\boldsymbol v_2=(-1,1)/\sqrt{2} $. Thus, at time $t$ the deviations from the synchronized state in the original basis and in the eigenbasis are related by the following equations: 
\begin{equation}
\begin{array}{l}
     y_{1,t}^{\xi}=\frac{1}{\sqrt{2}}\left(y_{1,t}^{\zeta}-y_{2,t}^{\zeta}\right), \hspace{20pt} y_{1,t}^{\zeta}=\frac{1}{\sqrt{2}}\left(y_{1,t}^{\xi}+y_{2,t}^{\xi}\right),\\
     y_{2,t}^{\xi}=\frac{1}{\sqrt{2}}\left(y_{1,t}^{\zeta}+y_{2,t}^{\zeta}\right), \hspace{20pt} y_{2,t}^{\zeta}=\frac{1}{\sqrt{2}}\left(-y_{1,t}^{\xi}+y_{2,t}^{\xi}\right).
  \end{array}
  \label{eq:transf1}
\end{equation}

The second column in Eq. \eqref{eq:transf1} helps to interpret what the eigenmodes represent. The first eigenmode is proportional to the sum of the deviations of the two agents. As it never converges to zero because $y_{1,t}^{\xi}$ and $y_{2,t}^{\xi}$ keep fluctuating (Figure \ref{fig:example_two_nodes}C), it represents the permanent phase shift in the dynamics due to the shock, also described in the literature as a perturbation that is ``parallel to the synchronization manifold''. The second eigenmode, which is ``transverse to the synchronization manifold'', is proportional to the difference of the deviations of the two agents from the synchronized state. As it eventually converges to zero, this suggests that the two agents always reach sychronization again after the shock. This can also be seen from the first column in Eq. \eqref{eq:transf1}: when $y_{2,t}^{\zeta}=0$, it is always $y_{1,t}^{\xi}=y_{2,t}^{\xi}=y_{1,t}^{\zeta}/\sqrt{2}$.

We can use this example to study analytically how other shocks would impact the dynamics. That is, we consider generic $u_1$ and $u_2$ hitting the agents at time $\tau$, so that $y_{1,\tau}^{\xi}=u_1$ and  $y_{2,\tau}^{\xi}=u_2$. From the second column of \eqref{eq:transf1} we can see that if $u_1=u_2=u$, $y_{2,\tau}^{\zeta}=0$, and so for all following time steps $t\geq \tau$, $y_{2,t}^{\zeta}=0$: The eigenmode transverse to the synchronization manifold never takes positive values.  In other words, because the shock is common, it does not cause any relative difference in phase between the two agents. On the contrary, the first eigenmode is maximal, $y_{1,\tau}^{\zeta}=\sqrt{2}u$: the effect of the shock is to only shift the phase, by the maximum amount. If $u>0$, the phase is shifted forward; if $u<0$, it is shifted backwards. Consider now the reverse case $u_1=-u_2=u$. Now $y_{1,\tau}^{\zeta}=0$ and $y_{2,\tau}^{\zeta}=\sqrt{2}u$. In this case, the transverse eigenmode is maximal, while the parallel eigenmode is null. So, after a transient in which the phases are different, they synchronize again, with no consequence of the shock in terms of permanent phase shift (this is the only case which does not imply a permanent phase shift).

\subsubsection{Six agents with two cliques of three nodes each}
\label{sec:six}

\begin{figure}[h!]
\centering
\includegraphics[width=1\textwidth]{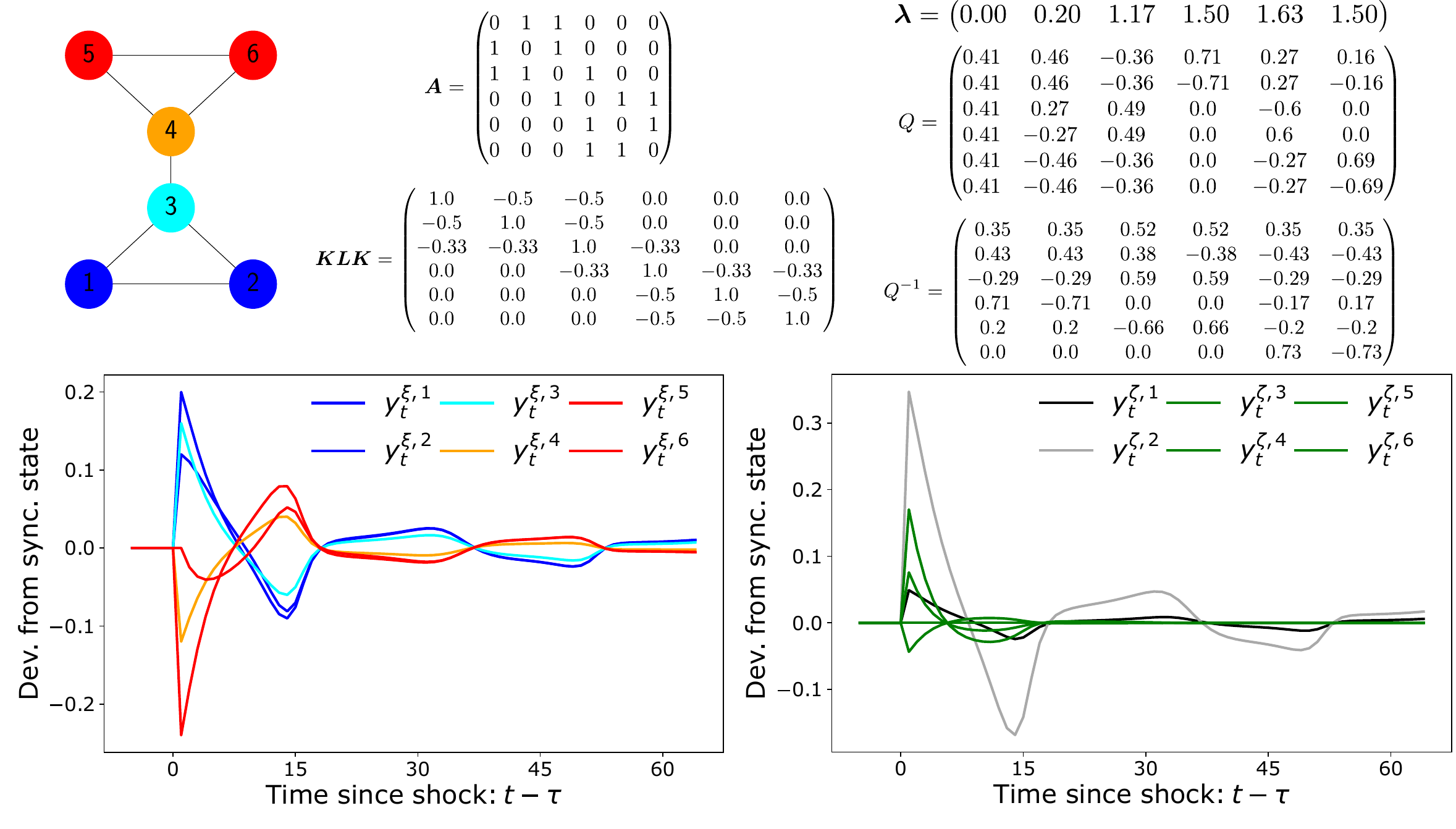}
\caption{Example dynamics in a network with six agents and two cliques of three nodes each. The top part of the figure shows the network, the adjacency matrix $\boldsymbol A$ and the normalized Laplacian $\boldsymbol K \boldsymbol L \boldsymbol K$. It also shows the eigenvalues $\boldsymbol \lambda$, the eigenvectors $\boldsymbol Q$ of the normalized Laplacian (each column of $\boldsymbol Q$ is the eigenvector corresponding to the eigenvalue shown above) and the inverse matrix of eigenvectors $\boldsymbol Q^{-1}$. The bottom part of the figure shows the evolution of the deviations in the original basis and in the eigenbasis.}
\label{fig:master_stab_N_6}
\end{figure}

To understand the effect of the network structure on shock propagation and synchronizing non-linear dynamics, we consider the network in Figure \ref{fig:master_stab_N_6}. This network is composed of two cliques of three nodes each, connected by a link between two of the nodes of the cliques. These two nodes are colored lighter to highlight the stronger connectivity to the other clique. This network can be thought of as a stylized model of six countries in two continents.

At time $\tau$, we apply the following shock vector:
\begin{equation}
\left(y^\xi_{1,\tau},y^\xi_{2,\tau},y^\xi_{3,\tau},y^\xi_{4,\tau},y^\xi_{5,\tau},y^\xi_{6,\tau}\right)=(0.20,0.12,0.16,-0.12,-0.24,0.00).
\end{equation}
This shock vector can be thought of as the combination of idiosyncratic shocks hitting all agents differently, combined with a positive shock hitting the clique with agents 1,2 and 3, and a negative shock hitting the other clique. In the eigenbasis, this shock vector corresponds to
\begin{equation}
\left(y^\zeta_{1,\tau},y^\zeta_{2,\tau},y^\zeta_{3,\tau},y^\zeta_{4,\tau},y^\zeta_{5,\tau},y^\zeta_{6,\tau}\right)=(0.05,  0.35,  0.00  ,  0.06, -0.08, -0.17).
\label{eq:ic_eigenbasis}
\end{equation}
It should be noted that the shock in the second eigenmode (0.35) is the largest one.

The evolution of the shocks is shown at the bottom of Figure \ref{fig:master_stab_N_6}. In the panel on the left, we show the shocks in the original basis, i.e. each line corresponds to an agent. It is clear that the dynamics vary across the two cliques, but they are very similar within each of the two cliques. The cyan and orange lines, corresponding to the two ``bridge'' agents, are the ones that show the closest dynamics to the ones in the other clique. Over time, all dynamics converge again to the synchronized state, with a permanent phase shift (not visible in the figure).

The evolution of the shocks in the eigenbasis is shown in the right panel. The black line is the eigenmode which is parallel to the synchronization manifold. The grey line corresponds to the second eigenmode, the one with largest initial value, and decays very slowly. The green lines correspond to the third to sixth eigenmodes, and decay very quickly. 

The interpretation for the eigenmodes is now clear. The second eigenmode corresponds to synchronization across cliques. Indeed, because $y_{2,t}^{\zeta}=0.43 y_{1,t}^{\xi} + 0.43 y_{2,t}^{\xi} + 0.38 y_{3,t}^{\xi} -0.38 y_{4,t}^{\xi}- 0.43 y_{5,t}^{\xi}- 0.43 y_{6,t}^{\xi}$, it is maximal in absolute value when shocks are positive in one clique and negative in the other one. This eigenmode takes time to decay, because $K = \epsilon \lambda_2 = 0.3 \cdot 0.2 = 0.06$ is relatively small and so, as shown in Figure \ref{fig:master_stability_function}, the largest Lyapunov exponent is not very negative ($\mu_1\approx-0.02$). This means that synchronization across cliques takes time, as the eigenmode decays by approximately 2\% every time step. The third to sixth eigenmodes correspond to synchronization within cliques, and decay quicker ($K=\epsilon\lambda_i$, $i=3,\ldots 6$ ranges from 0.35 to 0.49, so the largest Lyapunov exponent $\mu_1$ ranges between -0.14 and -0.18). This indicates that synchronization within cliques happens much quicker.

This phenomenon is general, because the second eigenvector of the Laplacian---the Fiedler vector---takes positive values in a cluster of nodes, and negative values in the other one. If there were $M$ cliques, or more loosely $M$ clusters of highly connected nodes weakly connected across one another, there would be $M-1$ small eigenvalues and then a large gap to the next eigenvalue. The eigenmodes corresponding to the small eigenvalues would indicate, as in the case above, synchronization across clusters. The remaining $N-M$ eigenvalues would be much larger, and the associated eigenmodes would correspond to synchronization within clusters.\footnote{There could also be a hierarchy of clusters. Imagine for example that there were two main clusters, and two sub-clusters within each cluster. Here, the Fiedler vector would divide between the two main clusters, but then the following eigenmode (corresponding to a larger eigenvalue) would correspond to synchronization between sub-clusters. There would be a time-scale separation so that synchronization would first occur within sub-clusters, then across sub-clusters within the same main cluster, and finally across main clusters.}

\subsubsection{Weighted directed networks}
\label{sec:supp_real_networks}

Due to the way in which we built the networks in the examples above, the analysis was performed for unweighted, undirected networks. However, nothing prevents us to apply it to the case of directed, weighted networks described by a generic connectivity matrix $\boldsymbol W$ (so that all rows are normalized to one): instead of calculating the eigenvalues of $\boldsymbol K \boldsymbol L \boldsymbol K$ one calculates those of $\boldsymbol I_N-\boldsymbol W$.

The only difference to the networks considered in the previous examples is that they have the convenient mathematical property that all eigenvalues of the normalized Laplacian are real, positive and bounded between zero and two. When considering a generic interaction network, the eigenvalues could instead be complex. Indeed, in the example considered in Section \ref{sec:real} we find that the eigenvalues have a very small imaginary part. Applying master stability analysis in the case of complex eigenvalues is almost identical to the case of real eigenvalues, except that one computes complex Floquet exponents (instead of real Lyapunov exponents) and then verifies if the complex values of $K=\epsilon \lambda_i$ are inside the stable region in the complex plane \citep{pecora1998master}. Here, for simplicity we just consider the real part of the eigenvalues and eigenvectors, since the imaginary part is small relative to the real part.

\section{International trade data}
\label{sec:supp_network_data}

In the international trade network interpretation of the model (Section \ref{sec:internationalmodel}), the interaction coefficients between countries represent the share of a country's output that is exported to another country. Importantly, the model also features ``self-loops'', representing the share of a country's output that remains within the same country. This internal demand, coming from domestic firms and households, is generally much higher than international demand. This makes it possible to set the problem as one of endogenous business cycles that originate within countries and synchronize internationally, as opposed to a problem in which business cycles originate \textit{because of} international linkages (as previously stressed, independent oscillations are a necessary ingredient to apply synchronization theory). Further to the importance of internal demand, another key feature of the model is that countries demand fixed proportions of imports from other countries. These two features of the model are crucial to determine which data should be used to calibrate the interaction network.

To have a consistent framework that includes information on internal demand, exports and imports, we turn to international input-output tables. Although we do not disaggregate our model at the industry-country level and do not distinguish between intermediate and final products, we prefer to use international input-output data over computing internal demand from raw import and export bilateral data and separately collected gross output data. Moreover, to test how good the assumption of fixed proportions of imports is, we need international input-output data that span as many decades as possible. 

For these reasons, the ideal dataset is a merge of the recently released Long-Run World Input-Output Database (LR-WIOD), covering the period 1965-2000 \citep{woltjer2021long}, with the established 2013 release of the WIOD, spanning the years 1995-2011 \citep{timmer2015illustrated}. Although the creators of these databases warn against using them together at a detailed industrial level  because of differences in their construction \citep{woltjer2021long}, they say that ``aggregate levels and trends are likely to be comparable''. Selecting the countries that are available in both datasets, we end up with 24 countries and a Rest of the World aggregate.

\begin{figure*}[h!]
\centering
\includegraphics[width=0.8\textwidth]{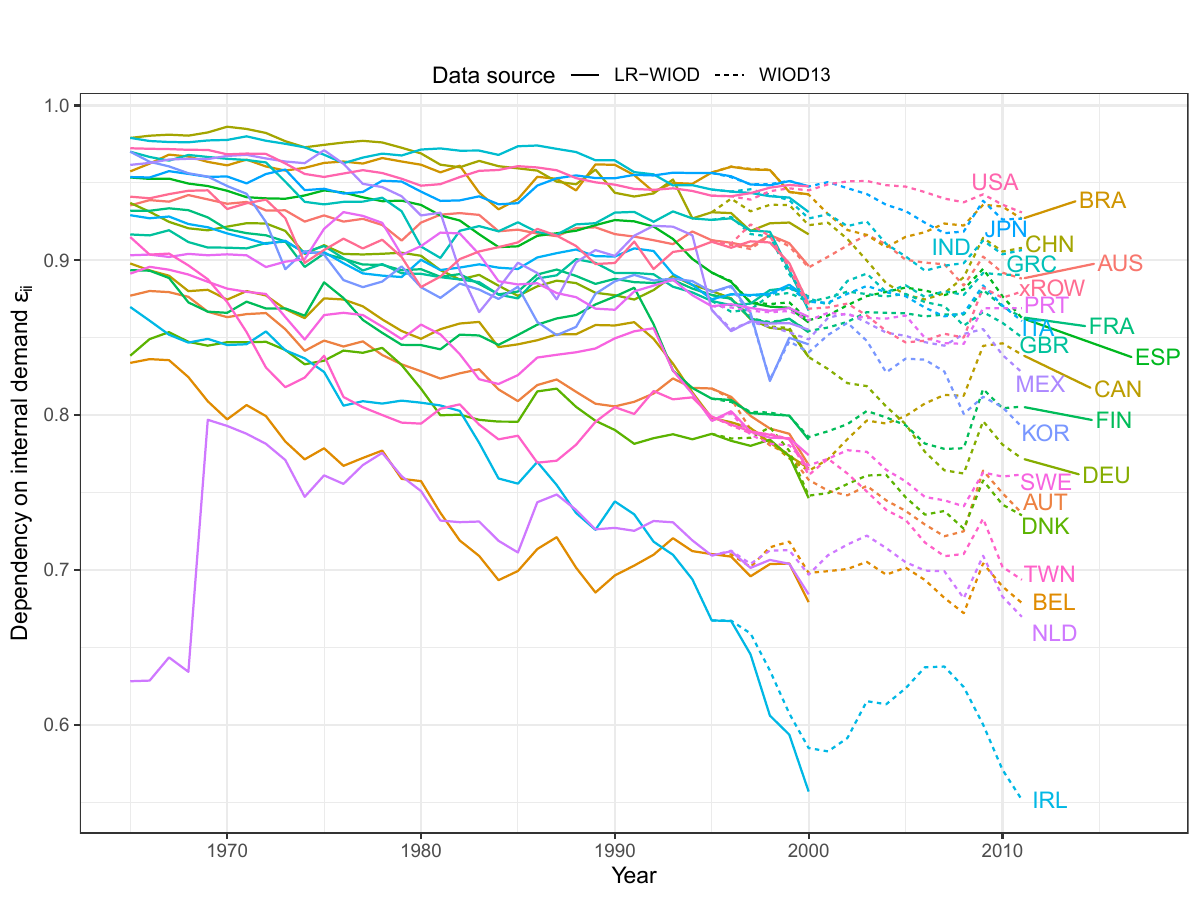}
\caption{Dependency on internal demand (self-loops), 1965-2011, for all countries in our sample. Solid lines are taken from LR-WIOD, whereas dashed lines are taken from WIOD 2013. }
\label{fig:compatibility_after_correction_edited}
\end{figure*}

Figure \ref{fig:compatibility_after_correction_edited} shows the dynamics of the share of internal demand for all the countries we consider. We show both data from LR-WIOD and data from WIOD 2013, including data from both datasets in the period 1995-2000. This figure has three take-away messages.

First, the trends in the overlapping period do indeed look similar across the two datasets. This justifies our choice to merge the two datasets and supports the hypothesis of \cite{woltjer2021long}. To perform the merging, we use LR-WIOD up to 1994 and WIOD 2013 from 1995 on, after adjusting the WIOD 2013 data so that the levels in 1995 match with those of LR-WIOD (note however that the levels were very close even without this transformation). Second, except for some irregular high-frequency movement, the share of internal demand changes slowly, generally decreasing. Similarly, we find that the fraction of bilateral exports over countries' output changes slowly over time (not shown), compared to business cycle frequency. We think that these results support our assumption of assuming fixed import shares as a first approximation. We will show that our results are robust to assuming time-varying interaction coefficients (Section \ref{sec:time-varying}). Third, the share of internal demand is almost always above 0.7, except for Ireland in recent years. This supports our view that business cycles originate from strategic complementarities of demand within countries, and then synchronize through international linkages, rather than originating from international demand.

\begin{figure*}[h!]
\centering
\includegraphics[width=1\textwidth]{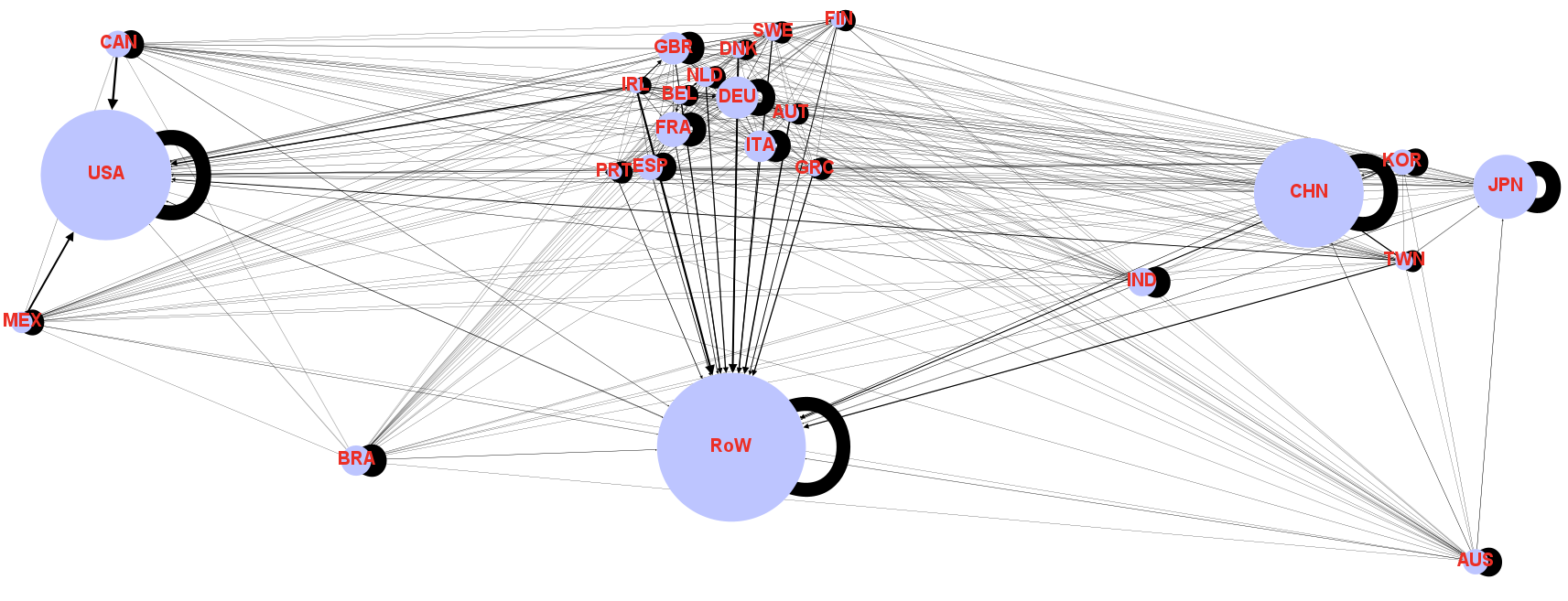}
\caption{Interaction network in 2011. The position of the node is equal to the countries' geographical centroid. Node size is proportional to countries' gross output, while link size is proportional to interaction coefficients $\epsilon_{ij}$ (for visualization purposes, we applied nonlinear transformations to both node and link size). RoW indicates the Rest of the World.}
\label{fig:network_visualization}
\end{figure*}

We show an example of the interaction network $\epsilon_{ij}$ in Figure \ref{fig:network_visualization}, for the most recent year available in our dataset. 

\section{Empirical application}
\label{sec:empirical_supp}

\subsection{Additional result on pairwise correlations}
\label{sec:pairwise_supp}

\begin{figure}[h!]
\centering
\includegraphics[width=0.5\textwidth]{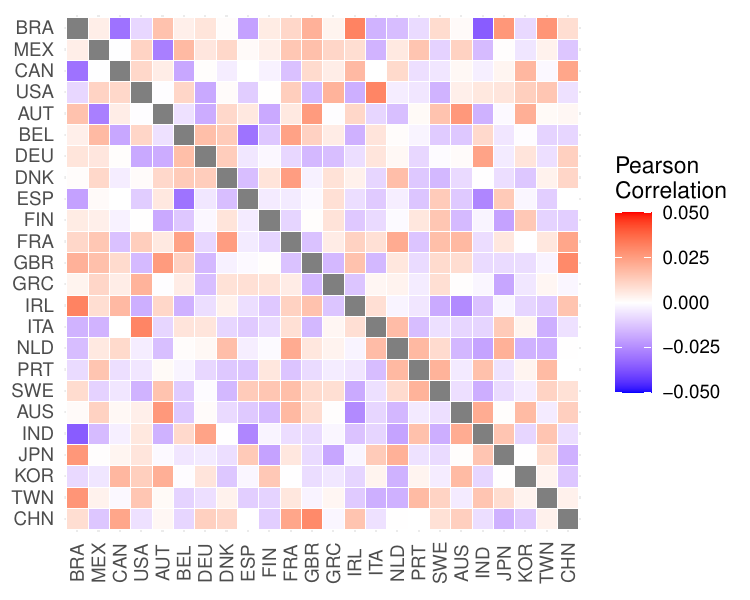}
\caption{Pairwise Pearson correlation coefficients between the dynamics in different countries when the model is parameterized as a node, averaged across 100 replications. }
\label{fig:baseline_pairwise_supp}
\end{figure}

Figure \ref{fig:baseline_pairwise_supp} shows that, when the model is parameterized as a node, it not only produces very low correlations, but there is also no clear pattern showing that pairs of countries with similar components of the Fiedler vector have higher pairwise correlation.

\subsection{Robustness}
\label{sec:robustness}

We now test if the main result is robust to alternative specifications.

\subsubsection{Shock persistence parameter}
\label{sec:persistence}

So far, we have used a value $\rho=0.3$ for the autocorrelation of idiosyncratic shocks. What happens when this persistence parameter varies?

\begin{figure}[h!]
\centering
\includegraphics[width=1\textwidth]{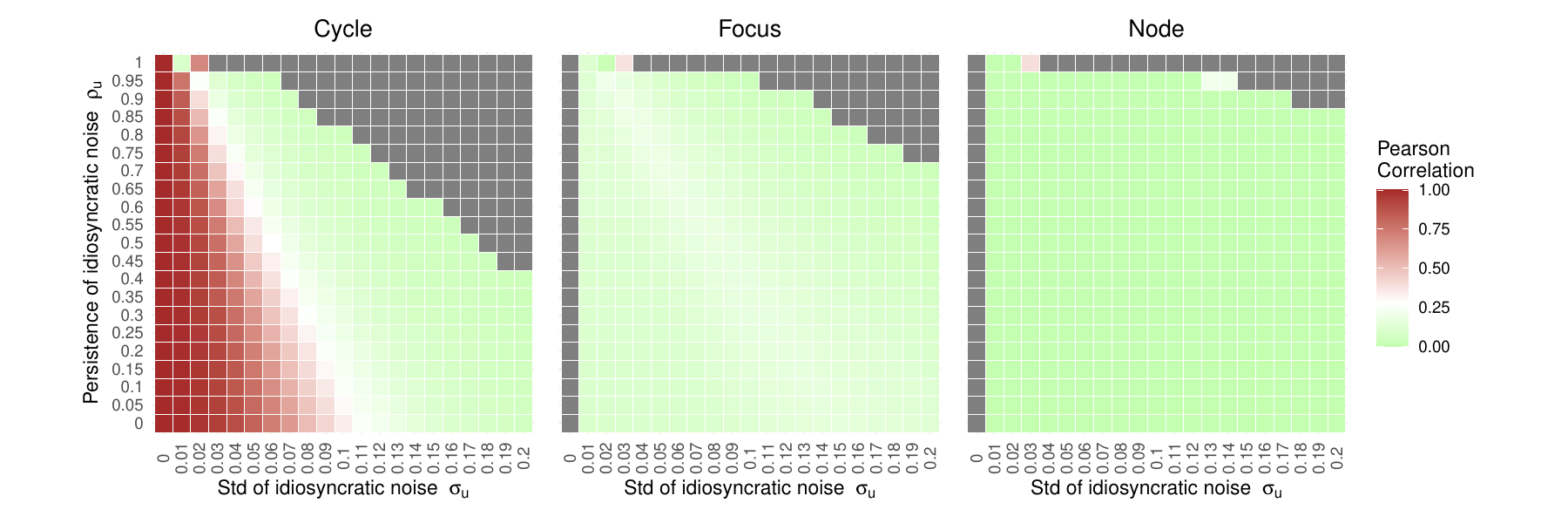}
\caption{Heatmaps showing the mean Pearson correlation coefficients across all country pairs and 20 simulation runs, for several values of the persistence parameter $\rho$ and the standard deviation of idiosyncratic shocks $\sigma$. The color bar is designed so that white corresponds to the mean of the empirical correlation of employment and value added. We show three heatmaps, for each of the main parameterization of the model.}
\label{fig:robustness_persistence}
\end{figure}

Figure \ref{fig:robustness_persistence} shows that, as we increase the persistence parameter, correlation coefficients become lower in the limit cycle case. This is not surprising, as higher persistence makes the shock processes lead the deterministic dynamics farther away from the values it would take in the absence of shocks. However, for all values of $\rho$, there is a value of $\sigma$ that makes the limit cycle model match the empirical level of comovement. By contrast, in the focus and node cases there is not a big effect of the persistence parameter, and in no case the model can produce correlations that are as high as in the data.

\subsubsection{Time-varying network}
\label{sec:time-varying}

From the description of the model in Section \ref{sec:internationalmodel}, we have been assuming that the interaction coefficients $\epsilon_{ij}$ between country $i$ and country $j$ are fixed over time. This was in part supported by the discussion of the data in Section \ref{sec:networkdata}, where we argued that the share of a country's output that is exported to other countries varies slowly over time, at lower frequencies than typical business cycle frequencies. Here, we relax the assumption of fixed interaction coefficients, and directly input to the model time-varying interaction coefficients $\epsilon_{ij,t}$, with $t=1965,\ldots,2011$, corresponding to the available data.

\begin{figure}[h!]
\centering
\includegraphics[width=0.6\textwidth]{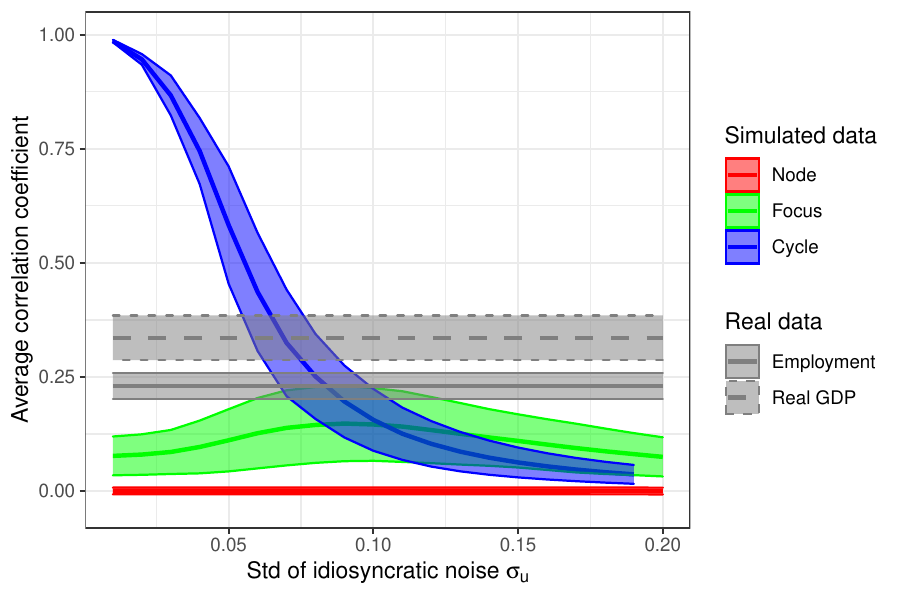}
\caption{Mean Pearson correlation across the decision variables $y$ of the countries in our sample as a function of the standard deviation of idiosyncratic shocks, in the model and in the data. The interpretation is the same as in Figure \ref{fig:baseline_correlations_average}.}
\label{fig:robustness_time_varying_network}
\end{figure}

As we can see in Figure \ref{fig:robustness_time_varying_network}, the results are virtually unchanged from Figure \ref{fig:baseline_correlations_average}, showing that indeed assuming fixed interaction coefficients is a good approximation for this problem.

\subsubsection{Other parameter values}
\label{sec:otherpars}

Next, we test if our results are also valid for parameters that are different from the baseline parameters indicated in Appendix \ref{sec:numerical_example}. To do so, still using the $F_1$ specification of the interaction function (Eq. \eqref{eq:F1}), we check for alternative values of strategic complementarities at the steady state by randomly picking the $\beta$ parameters. We restrict the parameter values so that they do not produce multiple equilibria and do not exit the stable region in Figure \ref{fig:stability_model} from below, as this would lead to a flip bifurcation and a limit cycle of period 2, which is unlikely to be relevant in economic analysis \citep{beaudry2015reviving}.

\begin{figure}[h!]
\centering
\includegraphics[width=1\textwidth]{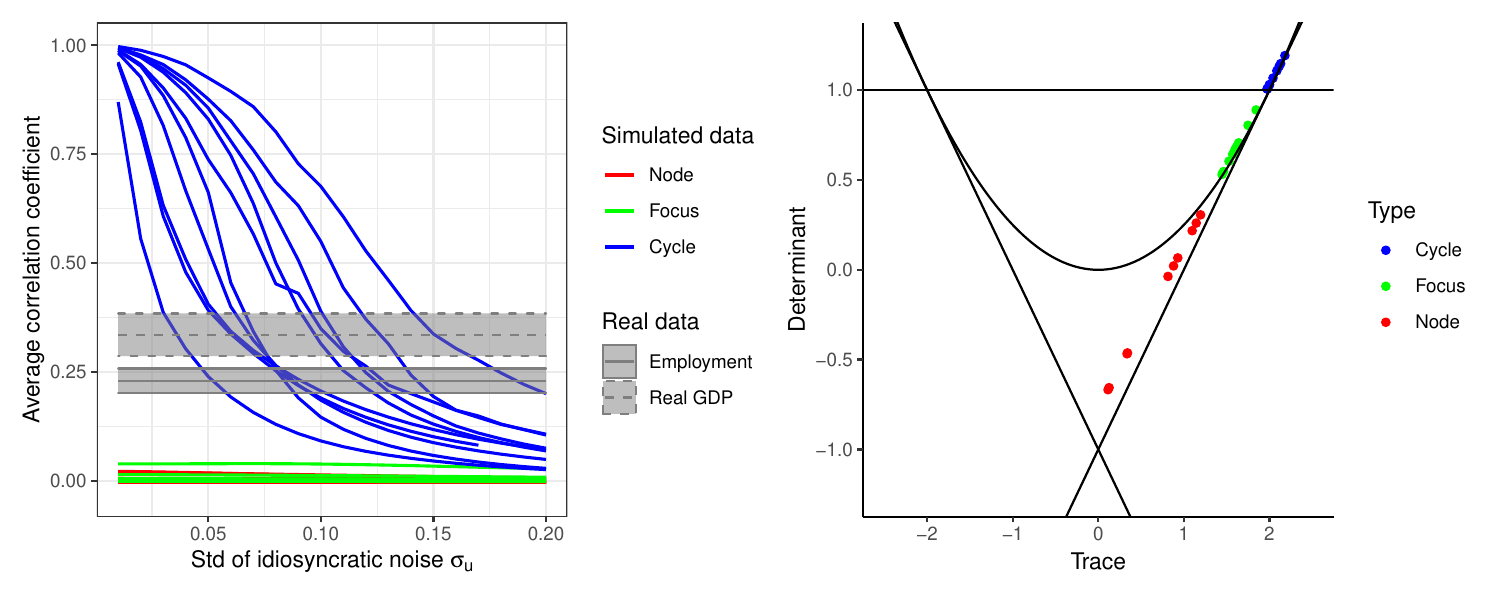}
\caption{Left: Same as in Figures \ref{fig:baseline_correlations_average} and \ref{fig:robustness_time_varying_network}, except that we only show results for many more parameter values than the baseline. We show the mean across 20 simulation runs (and not the error bands) and 10 combinations of the $\beta$ parameters for each of the cycle, focus and node cases. Right: same diagram as in Figure \ref{fig:stability_model}, for the same combinations of the $\beta$ parameters shown on the left. }
\label{fig:robustness_other_pars}
\end{figure}

As we show in Figure \ref{fig:robustness_other_pars} (left panel), the key results are robust to alternative parameter specifications, as only in the limit cycle case the model can produce correlations that are as high as in the data. Depending on the specific parameterization, the general level of comovement in the limit cycle case can be higher or lower than the baseline for a given strength of idiosyncratic shocks. This is because, depending on the parameters, limit cycle fluctuations have different amplitude, and so can be more or less perturbed by idiosyncratic shocks with a given standard deviation. It is also possible to note that we do not see a pattern for the focus case that is similar to Figure \ref{fig:baseline_correlations_average}, as in this case the maximal correlation is 0.04, much smaller than 0.15 as in Section \ref{sec:mainresult}. This is because, in that case, we selected the parameters so that the focus was very close to the boundary of the unstable region, and so oscillations driven by exogenous shocks would die out very slowly. This is not the case for the parameter combinations that we select here in the focus case, which are well within the stable region (Figure \ref{fig:robustness_other_pars}, right panel).

\subsubsection{Logistic specification of the interaction function}
\label{sec:logisticF}

Finally, we also vary the function $F$ mediating the effect of interactions, using the alternative logistic specification in Eq. \eqref{eq:F2}. In this case, there is only one free parameter, namely the slope at the steady state $\beta$, proportional to the strength of strategic complementarities. Thus, we simply consider 10 equally spaced values for $\beta$ for each of the cycle, focus and node regimes.

\begin{figure}[h!]
\centering
\includegraphics[width=1\textwidth]{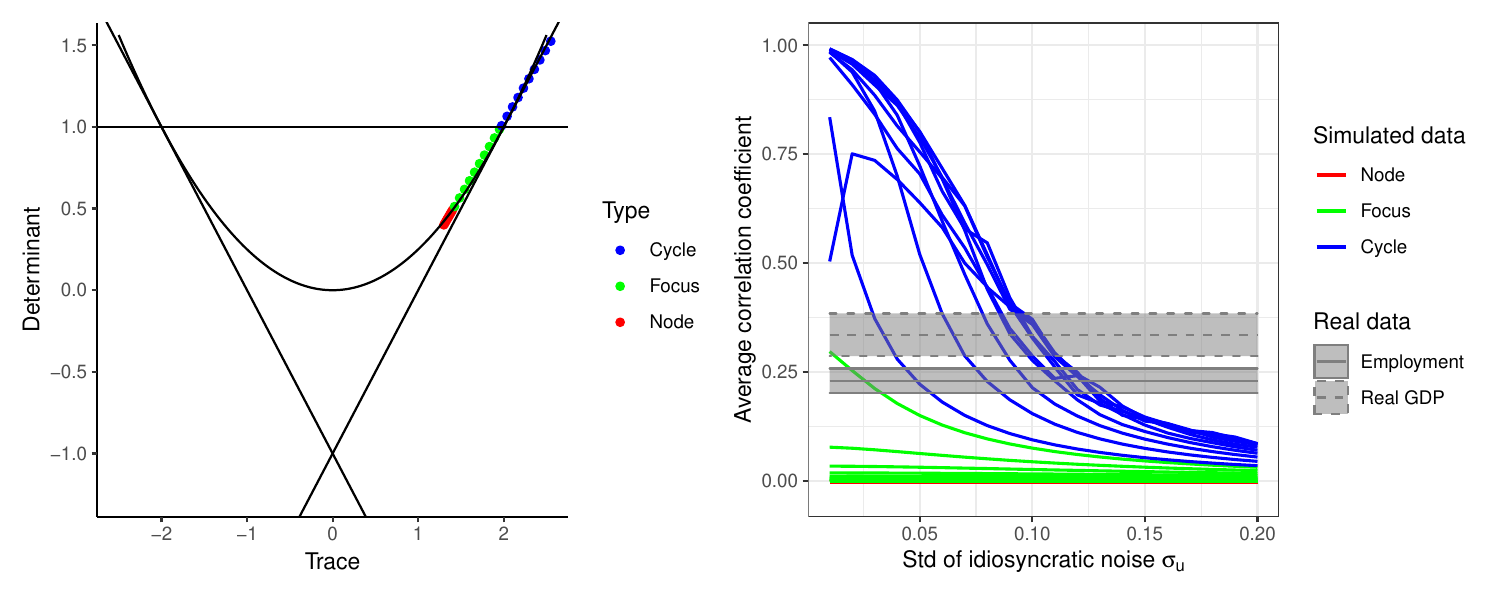}
\caption{Left: Same as Figure \ref{fig:robustness_other_pars}, except that for each of the cycle, focus and node cases we select 10 values of the $\beta$ parameters, representing the slope of the logistic function at the steady state. }
\label{fig:robustness_logistic_function}
\end{figure}

Figure \ref{fig:robustness_logistic_function} is very similar to Figure \ref{fig:robustness_other_pars}, mainly suggesting that the specification of the function $F$ as a quartic is not particularly restrictive. It is also interesting to note that, for the focus parameterization that is very close to the Hopf bifurcation, and for small levels of noise, comovement in the model can match the empirical comovement of employment, although it is still too small to match the empirical comovement of GDP.

\subsubsection{Bifurcation parameters unrelated to strategic complementarities}
\label{sec:bifurcation_no_compl}

Throughout the paper, by varying the $\beta$ parameters, we have used the strength of strategic complementarities at the steady state as a bifurcation parameter: as strategic complementarities increase, the steady state moves from being a node, to being a focus, to eventually become unstable and at the center of a limit cycle or chaotic attractor. 

Our choice of focusing on strategic complementarities as a bifurcation parameter leads to a possible criticism: what if it is the strength of strategic complementarities, rather than the type of dynamics, that causes higher comovement? 

To address this issue, we fix the parameterization of the limit cycle considered as a baseline, and consider two alternative set of parameters leading to the steady state being a node and a focus. For these two alternative parameter sets, we keep the $\beta$ parameters fixed (and so the regime of strategic complementarities), and vary instead $\alpha_1$, $\alpha_2$ and $\delta$ to get different stability properties of the steady state. Specifically, we use:
\begin{itemize}
\item Node: $\alpha_1=-0.11$, $\alpha_2=0.2$, $\delta=0.7$
\item Focus: $\alpha_1=-0.04$, $\alpha_2=0.2$, $\delta=0.1$
\end{itemize}
Thus, every difference in the resulting comovement must be due to the different dynamical regimes, rather than to the strength of strategic complementarities. 

\begin{figure}[h!]
\centering
\includegraphics[width=0.6\textwidth]{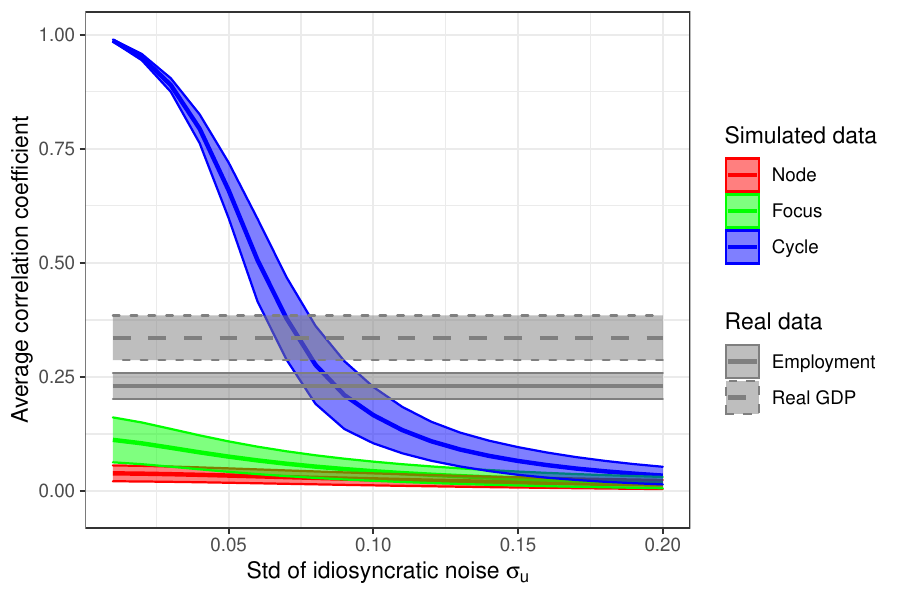}
\caption{Same as figures above, but in this case the parameters that lead to the node and focus preserve strategic complementarities, as detailed in Appendix \ref{sec:bifurcation_no_compl}. }
\label{fig:baseline_correlations_bifurcation_no_F}
\end{figure}

Figure \ref{fig:baseline_correlations_bifurcation_no_F} shows that the results are very similar to the previous figures. Thus, it is the type of dynamics, not the strength of strategic complementarities, that leads to different levels of comovement.

\subsubsection{Parameters heterogeneous across countries}
\label{sec:heterogeneous_alpha}

Throughout the paper we assumed that all nodes/countries are described by the same parameters. This was mostly out of analytical convenience, as the theory of complete synchronization developed in Appendix \ref{sec:masterstabilityanalysis} is cleanest in this case (although it can accommodate limited parameter heterogeneity, see \citealt{pikovsky2003synchronization}). However, nothing prevents us from testing the effect of country heterogeneity through simulation.

To do so, we select the $\alpha_1$ and $\alpha_2$ parameters, and allow them to vary across countries. Specifically, before we start the simulations we draw a random number from a normal distribution with mean zero and standard deviation 10\% of the baseline parameter values (that is, those listed in Appendix \ref{sec:numerical_example}), and add these random numbers to the baseline parameters.

\begin{figure}[h!]
\centering
\includegraphics[width=0.6\textwidth]{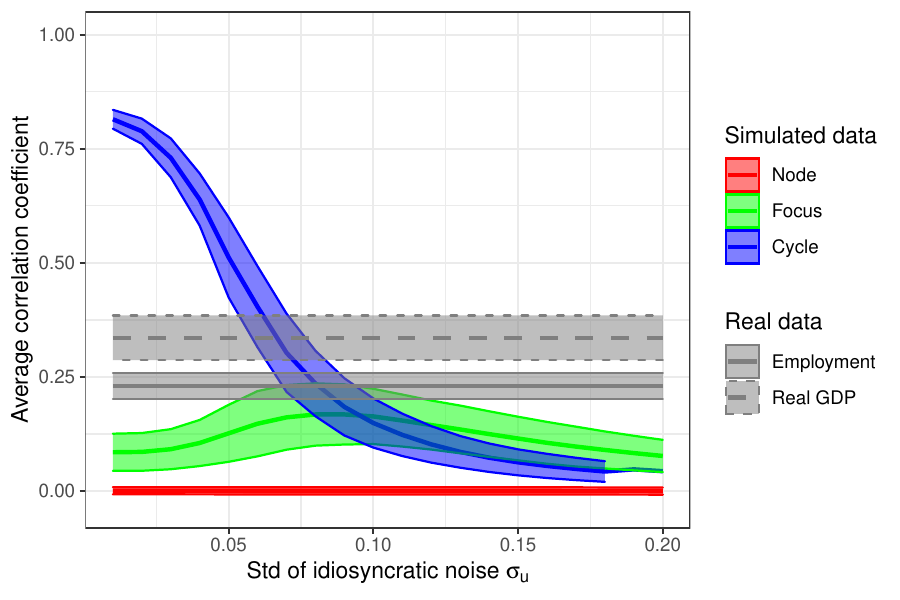}
\caption{Same as figures above, but in this case the parameters $\alpha_1$ and $\alpha_2$ are heterogeneous across countries. See Appendix \ref{sec:heterogeneous_alpha} for more details. }
\label{fig:baseline_correlations_heterogeneous_alpha}
\end{figure}

We report results in Figure \ref{fig:baseline_correlations_heterogeneous_alpha}. This is almost indistinguishable from Figure \ref{fig:baseline_correlations_average}, suggesting that heterogeneity of parameters plays a limited role, at least when it is moderate. 

\subsubsection{Common shocks}
\label{sec:common_shocks}

Trivially, if the standard deviation of common shocks is large compared to the standard deviation of idiosyncratic shocks, all dynamical regimes (cycles, focus and node) would yield very similar level of comovement. We do not think that this situation is interesting, nor relevant: really global events such as the Covid-19 pandemic are rare, and so country-specific shocks should be more prevalent. 

Therefore, we set the standard deviation of idiosyncratic shocks $\sigma_u$ to 0.08, a value at which the limit cycle regime matches the empirical level of comovement. Next, we vary the standard deviation of common shocks $\sigma_v$ from 0 to 0.04, following the assumption that common shocks can be at most half as large as country-specific shocks.

\begin{figure}[h!]
\centering
\includegraphics[width=0.6\textwidth]{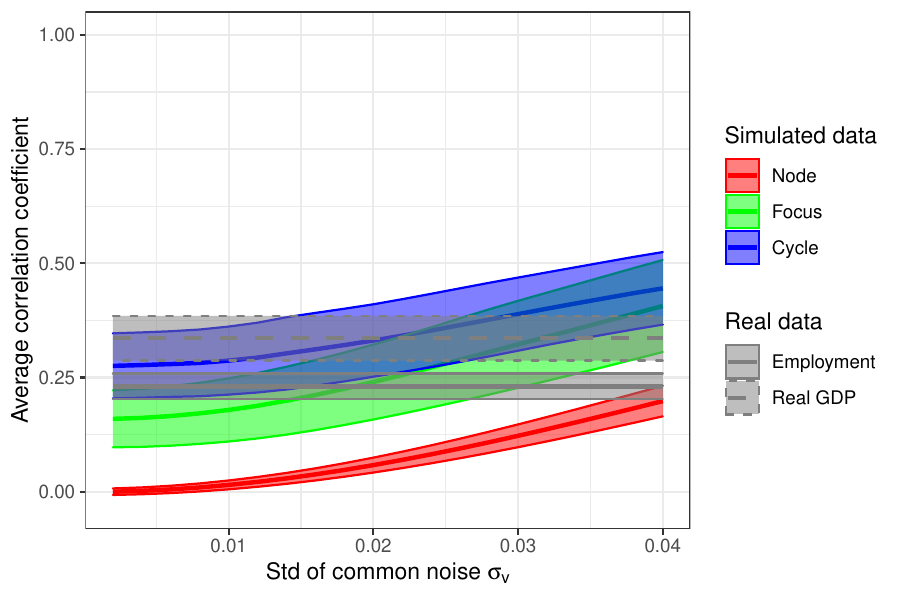}
\caption{Same as figures above, but in this case we vary the standard deviation of common shocks, for fixed standard deviation of idiosyncratic shocks. See Appendix \ref{sec:common_shocks} for more details. }
\label{fig:baseline_correlations_common_shocks}
\end{figure}

Figure \ref{fig:baseline_correlations_common_shocks} shows the results. We find that for only the largest values of $\sigma_v$ the node regime can match the level of comovement of employment, while it cannot match the level of comovement of GDP for any level of common shocks. Instead, the focus regime matches both the level of comovement of employment and GDP for sufficiently large values of $\sigma_v$ in this range.

These results confirm that common shocks are a powerful method to obtain high comovement. However, we emphasize that this is a great simplification, as in our view modeling common shocks as a common AR(1) process is a very strong assumption---properly modeling common shocks requires using a more structural model.

\clearpage

% Bibliography
\bibliographystyle{C:/Users/marco/Dropbox/Latex/econ}
\bibliography{C:/Users/marco/Dropbox/Latex/bibliografia}

\end{document}